\documentclass{elsart}
\usepackage{bm,txfonts}
\usepackage{amsfonts,float}
\usepackage{epsfig}
\newcommand{\beq}{\begin{equation}}
\newcommand{\eeq}{\end{equation}}
\newcommand{\beqa}{\begin{eqnarray}}
\newcommand{\eeqa}{\end{eqnarray}}
\newcommand{\nn}{\nonumber \\}
\newcommand {\np}[1]{ {\mbox{\textrm{:}}{#1}{\textrm{:}}} }
\newcommand\Dot[3]{{#1}\hspace{-#3em}\raisebox{#2em}{$\scriptscriptstyle\bullet$}}

\def \a {\underline{\alpha}}

\def \ch {\mathrm{ch}}
\def \D {\Delta}
\def \df {\stackrel{\mathrm{def}}{=} \ }
\def \e {\mathrm{e}}
\def \el {\mathrm{el}}
\def \eps {\varepsilon}
\def \g {\underline{\gamma}}
\def \inn {\mathrm{in}}
\def \max {\mathrm{max}}
\def \J {\mathcal J}
\def \k {\kappa}
\def \l {\lambda}
\def \L {\underline{\Lambda}}

\def \o {\underline{\omega}}
\def \out {\mathrm{out}}
\def \s {\sigma}
\def \t {\tau}
\def \qh {\mathrm{qh}}
\def \q {\underline{q}}
\def \Q {\underline{Q}}
\def \qp {\mathrm{qp}}
\def \z {\zeta}
\def \PF {\mathrm{PF}}
\def \Im {\mathrm{Im} \, }
\def \Re {\mathrm{Re} \, }
\def \mod {\ \mathrm{mod} \ }
\def \H {{\mathcal H}}
\def \Sc {{\Dot S{.74}{.48}\hspace{0.1cm}}}
\def \Qc {{\Dot Q{.74}{.38}}}
\def \uu {{\widehat{u(1)}}}
\def \C {{\mathbb C}}
\def \N {{\mathbb N}}
\def \R {{\mathbb R}}
\def \Z {{\mathbb Z}}
\def \V {\mathcal{V}}

\begin{document}
\begin{frontmatter}
\title{A universal conformal field theory approach to the chiral 
persistent currents in the mesoscopic \\ fractional quantum Hall states}
\author{Lachezar S. Georgiev}
\ead{lgeorg@inrne.bas.bg}
\address{Institute for Nuclear Research and
Nuclear Energy \\
 Tsarigradsko Chaussee 72,  1784 Sofia, BULGARIA}
\begin{abstract}
We propose a general and compact scheme for the computation of the
periods and amplitudes of the chiral persistent currents, magnetizations
and magnetic susceptibilities 
in mesoscopic fractional quantum Hall disk samples threaded by 
Aharonov--Bohm magnetic field. 
This universal approach uses the effective conformal field theory for 
the edge states in the quantum Hall effect 
to derive explicit formulas for the corresponding partition functions
in presence of flux. 
We point out the crucial role of a special invariance condition 
 for the  partition function, following from the Bloch--Byers--Yang 
theorem, which represents the Laughlin spectral flow.
As an example  we apply this procedure to the $\Z_k$ parafermion Hall 
states  and show that they have universal non-Fermi liquid behavior 
without anomalous oscillations. For the analysis of the high-temperature 
asymptotics of the persistent currents in the parafermion states
we derive the modular $S$-matrices constructed from the 
$S$ matrices for the $u(1)$ sector and that for the neutral 
 parafermion sector which is realized as a diagonal affine coset.
\end{abstract}
\end{frontmatter}
\section{Introduction}
Mesoscopic physics has recently become a very intensive area of research
because of the  exciting opportunities it offers to observe fundamental  
implications of quantum mechanics in experimentally 
realizable two-dimensional samples. 
One important characteristics of the  mesoscopic phenomena is that 
the electrons  are coherent which implies that the 
sizes should be extremely small, the samples extremely pure and with 
 high mobility and the temperatures very low. 

According to a famous Bloch theorem
the free energy of a conducting ring threaded by magnetic field
 is a periodic function of the magnetic flux through the
ring with period one flux quantum $\phi_0=h/e=1$.
The flux dependence of the free energy
\beq\label{F}
F(T,\phi)=-k_B T \ln(Z(T,\phi)),
\eeq
where $Z(T,\phi)$ is the partition
function  gives rise to  an equilibrium current
\beq\label{I}
I(T,\phi)=-\left(\frac{e}{h}\right) 
\frac{\partial  F(T,\phi) }{ \partial \phi} ,
\eeq
called a
\textit{persistent current},
flowing along the ring without dissipation.
These oscillating currents have  universal amplitudes
and temperature dependences and have been  observed in mesoscopic 
rings \cite{pers-exp}, where the length of the ring is smaller 
than the coherence length at very low temperatures.

The fractional quantum Hall (FQH) states could be generically interpreted as 
mesoscopic rings because the only delocalized states live on the 
edge of the two-di\-men\-sio\-nal disk sample. 
The universality classes of the FQH
systems have been successfully  described by the effective field
theories for the edge states in the thermodynamic scaling
limit \cite{fro-stu-thi}. These turned out to be
topological field theories in $2+1$ dimensional space-time which are
equivalent to conformal field theories  (CFT) on the $1+1$
dimensional space border, provided the bulk excitations are suppressed
by a nonzero energy gap. 

Most of the computations of the persistent currents in
strongly correlated mesoscopic systems, including in the FQH  
 regime,  so far have been
based mainly on the Fermi or Luttinger liquid picture 
\cite{geller-loss-kircz,geller-loss}
because  a more general approach to these quantities was missing.
In this paper we shall formulate a general and compact
scheme for the computation of the chiral persistent currents 
in mesoscopic FQH disk states  based on the
rational CFT on  the edge.
 This scheme has become very important after
recent experiments \cite{grayson} showed that the (multicomponent) 
Luttinger liquid theory is irrelevant for more general filling factors such as
those in the principle Jain series. We stress that the
rational CFT description which is the right strategy for classifying the
FQH universality  classes is also the right starting point for the 
computation of the persistent currents in the FQH states.
The main advantage of this approach  is that the CFT partition function
which is used as a thermodynamical potential 
can be computed analytically even when Aharonov--Bohm (AB) magnetic field 
is threading the sample. This fact has already been used in 
Refs.~\cite{5-2} and
\cite{PRB-PF_k} for the computation of the persistent currents, 
magnetizations  and the corresponding magnetic susceptibilities 
with the intuitive explanation that because the transformation
$\z\to\z+\t$ represents the addition of one flux quantum  its 
generalization $\z\to\z+\phi\t$ should be interpreted as 
adding flux $\phi$ in units 
$\phi_0=1$. We derive this result in Sect.~\ref{sec:AB} which was one 
of the motivations for the present paper.
The fact that the magnetic field is of AB
type (i.e., an infinitely long and thin solenoid) is a significant 
 simplification because the AB magnetic field is zero at the 
edge so that it does not break the conformal invariance. On the other 
hand this limitation is not completely restrictive  because the edge 
of a FQH system is so  thin that the free energy depends only on the 
flux encircled by the edge and not on the magnetic field alone.
Adding AB flux to the FQH system naturally introduces twisted 
boundary conditions for the field operators of the electron and  
quasiparticles and also modifies the effective Hamiltonian. 
We show that the ultimate effect of this twisting on the
partition function is that it is multiplied by a universal exponential 
factor and that the modular parameter $\z$ is shifted by an amount 
proportional to the modular parameter $\t$ where the coefficient of 
proportionality coincides with the AB flux $\phi$.
As an illustration of the general CFT approach to the chiral 
persistent currents we are computing these currents for the parafermion 
FQH states using the partition functions developed in \cite{NPB2001}.

The  persistent currents in the FQH regime typically contain  large 
non-mesoscopic components \cite{oscillate}, slowly varying with
the magnetic field, which cannot be described within the CFT framework
 since there
are contributions from states laying deep below the Fermi level.
The edge states effective CFT is relevant only for the computation of
the much smaller oscillating part of the persistent currents which
could nevertheless be tested directly in SQUID detectors \cite{pers-exp} 
using the Josephson effect.
While the persistent currents naturally express the electric properties 
of the system, it turns out that they could also give important information
about the neutral structure of the effective field theory for the edge
states \cite{geller-loss-kircz,ino}.
As we shall demonstrate in Sect.~\ref{sec:low-T} and
Sect.~\ref{sec:high-T} the low- and high- temperature decays of
the persistent current's amplitudes crucially depend on the neutral
properties  of the edge states. Our analysis reveals two different 
universal mechanisms for the thermal reduction of the persistent currents: 
at low temperatures the thermal activation of 
quasiparticle--quasihole pairs leads to a logarithmic decay of the 
amplitudes  while at high temperatures the reduction is due to thermal 
decoherence and the decay is exponential.

 The rest of this paper is organized as follows: 
in Sect.~\ref{sec:part} we review the rational CFT approach to 
the FQH states, describe in brief  the terminology of the chiral quantum Hall 
lattices of topological charges which is necessary for the understanding of
 the general classification of FQH universality classes.
In Sect.~\ref{sec:AB} we show how  the CFT partition function is modified
by the AB flux starting from  the twisting of the electron field and 
deriving explicitly the modification of the $\uu$ current and the 
Hamiltonian. 
In Sect.~\ref{sec:CZ} we discuss the Laughlin spectral flow, the role
of the Cappelli--Zemba factors for restoring the  invariance of the 
partition function under this flow and prove that the CFT 
partition function with AB flux is invariant under $\phi\to \phi+1$
as it should be according to the Bloch--Byers--Yang theorem.
In Sect.~\ref{sec:Z_low-T} we derive the low-temperature asymptotics 
of the CFT partition function, which is universal. 
This asymptotics is then used in 
Sect.~\ref{sec:low-T} to derive the low-temperature asymptotics 
of the persistent currents amplitudes which have a universal form and is 
expected to be valid for  all FQH states.
In Sect.~\ref{sec:Z_high-T} we show how to use the modular $S$ 
transformation in a rational CFT to analyze the high-temperature 
asymptotics of the partition functions and the persistent currents.
In Sect.~\ref{sec:PF} we summarize the results from Ref.~\cite{NPB2001} 
about the partition functions for the edge states
 of the parafermion (PF) FQH states for $k=2,3$ and $4$. 
We derive analytically in Sect.~\ref{sec:S} the complete $S$ matrix for 
the parafermion states
as well as that for the neutral part which is realized as a diagonal 
affine coset.
 Then we investigate the persistent currents for fixed temperature
 when the flux is varied within  one period 
 as well as the temperature decay of the amplitude.
 In  Sect.~\ref{sec:high-T} the high-temperature asymptotics
for the parafermion FQH states  is computed 
using the analytic $S$ matrices derived  in Sect.~\ref{sec:S}. 
 Some technical details are put in several appendices.
\section{Chiral partition function for the FQH states 
with Aharonov--Bohm flux}
\label{sec:part}
\subsection{CFT for the edge states: labeling  the FQH universality 
classes}
Due to  the existence  of a nonzero energy gap in the
FQH effect excitations spectrum the electron system behaves as an 
incompressible electron liquid at low energies. The effective field theory
in the scaling thermodynamic limit, which essentially 
describes the low-energy excitations,  turns out to be a topological 
relativistic quantum field theory in 2+1 dimensions 
with a Chern--Simons action. It is well known that these theories are 
equivalent to CFTs on the 1+1 dimensional edge.  Therefore, it is 
believed that the CFT could capture all universal properties of the 
FQH systems and thus allow us to make a complete classification 
of the QH universality classes.

The electric current on the edge, where the coordinates are 
denoted by $\{x_\alpha\}$ for $\alpha=1,2$ and $\{ A_\alpha\}$ is the 
restriction of the vector potential to the edge,  is anomalous \cite{fro2000}
\[
\partial_\alpha j^\alpha=\frac{1}{2}\sigma_H \epsilon_{\alpha\beta}
F_{\alpha\beta},
\quad \sigma_H=\frac{e^2}{h} \nu_H, \quad
F_{\alpha\beta}=\partial_\alpha A_\beta - \partial_\beta A_\alpha,
\]
i.e., it violates the $u(1)$
invariance on the edge, however, this is exactly compensated by the 
violation of the $u(1)$ invariance in the bulk so that the sum of the bulk 
and edge current is invariant. 
This compensation of bulk and edge anomalies 
is the origin of the so called holographic principle
\cite{fro2000}
which in this case can be formulated as the correspondence between the 
quasiparticle spectra on the edge and in the bulk.
The $u(1)$ anomaly mentioned above  is the basic mechanism 
through which the Hall current is transmitted between the two edges of 
the real FQH samples.
\subsection{The $\widehat{u(1)}\times Vir$ symmetry}
The chiral CFT describing the edge of a FQH system always contains 
a separate $\widehat{u(1)}$ and Virasoro ($Vir$) symmetries. The first one 
describes the electric charge $Q$, or equivalently the magnetic flux
$\phi$ properties due to the  fundamental FQH relation \cite{fro-stu-thi}
\beq\label{charge-flux}
Q=\nu_H \, \phi,
\eeq
($e=1$) while the second corresponds to the angular momentum. Note that
due to the Sugawara formula  
the $\uu$ sector contributes to the angular momentum as well so 
by $Vir$
in the  relative $\widehat{u(1)}\times Vir$ decomposition  we mean
the Virasoro contribution from the neutral sector only. As we shall see 
in Sect.~\ref{sec:q_el}
the neutral $Vir$ sector must be present for all FQH states with 
filling factors whose numerators are bigger than $1$.

The \textit{chiral partition function} $Z(\t,\z)$ corresponding to
a disk FQH sample with a single edge  
computed within the effective CFT for the edge states
\beq\label{Z_chi}
{Z}(\t,\z) = \mathop{\mathrm{tr}}_{\ \ \H \ } q^{L_0-c/24}\,
\exp\left(2\pi i \z J_0^\mathrm{el}\right) =
\sum_{\l=1}^\mathcal{N} \chi_{\l}(\t,\z), 
\eeq
is the \textit{linear} sum of all topologically distinct chiral CFT 
characters \cite{PRB-PF_k} 
\beq\label{chi_l}
\chi_{\l}(\t,\z) =  \mathop{\mathrm{tr}}_{\ \ \H_{\l} \ }
q^{L_0-c/24}\,  \exp\left(2\pi i \z J_0^\mathrm{el}\right)\ \  
\mathrm{where} \ \
\H=\mathop{\oplus}_{\lambda=1}^\mathcal{N} \H_\lambda.
\eeq
The number $\mathcal{N}$ in Eqs.~(\ref{Z_chi}), called the 
\textit{topological order} \cite{wen-top}, (the number of topologically 
inequivalent quasiparticles with electric charge \cite{fro2000} 
$0\leq Q_{\mathrm{el}} <1$)  
is one of the most important characteristics of the FQH universality classes.
The Hilbert space of the edge states $\H$, over which the traces in 
Eqs.~(\ref{Z_chi}) and (\ref{chi_l}) are taken, is the direct sum of all 
independent (i.e., irreducible, topologically inequivalent)
representations  $\H_\lambda$ of the chiral algebra which are closed 
under fusion (see the Introduction of Ref.~\cite{gaps} for a short 
description of the chiral algebra terminology in the FQH effect context).
The effective Hamiltonian for the edge states in the thermodynamic limit
is given by \cite{cz}
\beq\label{H_CFT}
H_{\mathrm{CFT}}= \frac{v_F}{R} \left( L_0 -\frac{c}{24}\right),
\quad L_0=\oint\frac{d\,z}{2\pi i} \ z\, T(z),
\quad z=\exp\left(\frac{v_F\, t-i\, x}{R}\right),
\eeq
where $x$ is the coordinate on the edge, $t$ is the 
imaginary time,  $v_F$ is the Fermi velocity on the edge, $R$ is its radius,
 $L_0$   is the zero mode of the Virasoro stress tensor $T(z)$ and $c$ is 
the central charge of the latter \cite{CFT-book}. 
The CFT Hamiltonian~(\ref{H_CFT}) captures the universal properties of 
realistic FQH systems  and  could  eventually give a qualitative description
of the Hilbert space structure of the low-lying excitations \cite{ctz3}, 
the energy spectrum  and even the (universal part of the) energy gap 
\cite{gaps}  of the FQH systems in the thermodynamic limit.
In particular, it  should not be  surprising 
 that  the CFT-based computations \cite{ino} of the persistent currents
in the Laughlin FQH states give the same results as those based on the 
Luttinger liquid picture \cite{geller-loss}.

The electric charge operator 
$J^\mathrm{el}_0$ in Eq.~(\ref{Z_chi}), defined as the space integral of the 
charge density  $J^\mathrm{el}(z)$ on the edge, 
which is proportional to the 
normalized $\widehat{u(1)}$ current 
\beq\label{J}
J(z)=i\,\partial\phi^{(c)}(z)=\sum_{n\in\Z}J_n\, z^{-n-1}, \quad 
\left[ J_n,J_m\right]=n\, \delta_{n+m,0},
\eeq
 has the following 
short-distance operator product expansion 
\cite{fro-stu-thi,CFT-book,cz,gaps} 
\beq\label{J_el}
J_0^\mathrm{el}=\oint\frac{d\,z}{2\pi i} \, J^\mathrm{el}(z),
\quad J^\mathrm{el}(z)=\sqrt{\nu_H} \, J(z),\quad 
J^\mathrm{el}(z)J^\mathrm{el}(w)\sim \frac{\nu_H}{(z-w)^2}.
\eeq
The modular parameters $q$ and $\z$ entering Eq.~(\ref{chi_l}) 
 are generically restricted in annulus samples for convergence by 
\[
q=\e^{2\pi i\t}, \quad \Im\t >0,
\quad \left( \Im\t\sim 1/k_B T\right) \quad \z\in \C
\]
and their real and imaginary parts are related to the inverse temperature, 
 chemical potential and Hall voltage \cite{cz}. 
For the disk FQH sample, however,
both  $\t$ and $\z$ have to be purely imaginary in order for
 the partition function Eq.~(\ref{Z_chi}) be real.
The physical meaning of $\Im\t$ and $\Im\z$ will be discussed in 
Sect.~\ref{sec:CZ}. 

Because the AB flux modifies only the charged  $\uu$ sector of the model
it would be convenient to artificially split the stress tensor into a 
charged and a neutral parts
\beq\label{Tc}
T(z)= T^{(c)}(z) + T^{(0)}(z),\quad \mathrm{where}\quad  
T^{(c)}(z)=\frac{1}{2}\,\np{J(z)^2}
\eeq
is the Sugawara stress tensor of $\uu$ and $T^{(0)}=T- T^{(c)}$ is its 
complement.
\subsection{Abelian rational CFTs: FQH states classification in terms 
of integer lattices of topological charges}
\label{sec:abelian}
The CFTs which naturally appear \cite{fro-stu-thi} in the FQH effect 
context are rational extension of the $\uu^N$ current algebra 
in terms of integer lattices of topological charges. A central result
in this approach is the complete A-D-E classification of the FQH 
universality classes in terms of 
abelian rational CFT associated with those lattices \cite{fro-stu-thi}.
We recall that an abelian theory is a rational conformal field theory 
with integer central charge $c=N$ and $\uu^N$ affine symmetry; 
the current algebra is generated by $N$ abelian currents 
$J^a (z)=i\partial_z \phi^a(z)$, extended by the
vertex operators $Y(\l,z)= \np{\exp(i\lambda_a\phi^a(z))}$,
whose ``topological charges'' $\lambda_a$ form  a $N$-dimensional 
lattice $\Gamma$. The lattices which appear in the quantum Hall effect 
must satisfy some specific physical conditions which are summarized 
bellow. 
Following Ref.\cite{fro-stu-thi}, we call {\it chiral quantum Hall lattice} 
an odd integral lattice to which a special {\it charge} vector $\Q$ from
 the dual lattice $\Gamma^*$ is associated. This vector 
 assigns to any lattice point $\q \in \Gamma$ the {\it electric charge} 
$(\q|\Q)$ of the corresponding edge excitation.
In particular, the lattice should contain an electron excitation $\q_\el$
with unit charge $(\q_\el|\Q)=- 1$. The norm 
$|\q|^2=(\q|\q)$ of the vectors $\q \in \Gamma$ gives twice the conformal 
dimensions, i.e. twice the spin of the corresponding excitation, while the 
scalar product $(\q|\q')$ is the relative statistics of the excitations 
$\q$ and $\q'$.

The charge vector $\Q$ satisfies the following conditions:
\begin{description}
\item[\ \ (i)] 
it is {\it primitive}, i.e., not a multiple of any other vector
$\q^*\in\Gamma^*$;
\item[ (ii)] 
it is related to $\nu_H$ by:
\beq
\label{0.3}
|\Q|^2=\nu_H=\frac{n_H}{d_H}, \quad n_H, d_H\in \N, \quad
\mathrm{gcd}(n_H,d_H)=1 ;
\eeq
\item[(iii)] it obeys the {\it charge--statistics relation} for boson/fermion
excitations:
\beq\label{0.3b}
(\Q|\q)=|\q|^2 \quad \mod \ 2\ , \qquad {\rm for \ any} \ \q\in\Gamma.
\eeq
\end{description}
We shall briefly recall the notion of  {\it maximal symmetry} 
\cite{fro-stu-thi}
since it covers the largest class of QH lattices.
A chiral  Hall lattice  is called
{\it maximally symmetric} if its neutral sublattice ($\Gamma_0 \perp \Q$)
coincides with the (internal symmetry) Witt sublattice
 $\Gamma_W \subset \Gamma$  (defined as the sublattice generated by
all vectors of square length 1 or 2) so that dim $\Gamma_W=$ dim 
$\Gamma  - 1$.
The Witt sublattice is always
of the type $A\oplus D \oplus E \oplus \Z$.
The maximally symmetric lattices, denoted in \cite{fro-stu-thi} by the
symbol $(L| \, {}^{\o}\Gamma_W)$, can be characterized in an appropriate basis
$\{ \q^i\}$ by the  following Gram matrix $G=(G^{ij})$ where 
$G^{ij}=(\q^i|\q^j)$ is the metrics of the lattice:
\beq
\label{MSQHL}
G =\left[
\begin{array}{c|c}
L &  \o \cr \hline
\o^T & C(\Gamma_W)
\end{array}
\right],  \quad \det(G) =\frac{1}{\nu_H}\det\left[C(\Gamma_W)\right];
\eeq
the charge vector is $Q=(1,0, \ldots, 0)$ in the corresponding dual basis.
Here $L\in\N$ is the {\it minimal relative angular 
momentum}\footnote{
In this case, it is equal to twice the conformal dimension of 
the electron operator, i.e. an odd integer.} 
that appears as
the minimal power of $(z_i - z_j)$ in the corresponding wave function;
$C(\Gamma_W)$ is the {\it Cartan matrix} for the Witt sublattice and
$\o$ is an {\it admissible weight} for $\Gamma_W$ restricted by the
condition  $(\o | \o)<L$ (see Eq. (5.6) in \cite{fro-stu-thi}).
Note that the Witt sublattice does not contribute to the filling
fraction $\nu_H$ since it is orthogonal to the charge vector $\Q$.
It is easy to find $\nu_H$ corresponding to the lattice with the Gram matrix 
(\ref{MSQHL}) using Eq.~(\ref{0.3}),
\beq
\label{nu-msqhl}
\nu_H = (\Q|\Q) = Q^T. G_\Gamma^{-1}.Q=\frac{1}{L-(\o|\o) }\ .
\eeq
The non-equivalent quasi-particles in the Hall state correspond to 
the irreducible representations of the extended algebra and are labelled by
the points $\underline{\l}\in\Gamma^*/\Gamma$ where 
$\underline{\l}$ is the quasiparticle's topological charge and 
$\Gamma^*$ is the  lattice dual to $\Gamma$. 
We recall that $\Gamma^*/ \Gamma$ is a 
finite abelian group of order $(\det\Gamma)$ whose multiplication law 
represents the fusion rules.
There are several important consequences  of the specific  structure of the 
Gram matrix (\ref{MSQHL}) and FQH effect specifics which we consider in more detail 
in the next subsections.
\subsection{The electron charge decomposition} 
\label{sec:q_el}
The Gram matrix (\ref{MSQHL}) implies that the  topological charge $ \q_\el $ 
of the electron should take the form \cite{fro-stu-thi}
\beq
\label{1.3}
\q_\el= - \frac{1}{\nu_H}\ \Q +\o  , 
\eeq
where $\Q$ is the lattice charge vector and  $\o$, which is the weight 
entering Eq.~(\ref{MSQHL}) i.e.,  an admissible weight of the
 Witt sublattice \cite{fro-stu-thi}, 
represents the neutral degrees of freedom 
of the electron because  $(\Q|\o)=0$. Then the electron field operator 
would be simply a tensor product of a $\uu$ vertex exponent of a chiral 
boson $\phi$, which is a simple current of the $\uu$ theory, and in 
general a simple current $\Phi_{\o}$ from the neutral part of the theory
\beq\label{el}
\psi_{\mathrm{el}}(z)= \np{\exp\left( -i\frac{1}{\sqrt{\nu_H}} 
\phi^{(c)}(z)\right)} 
\otimes \ \Phi_{\o}(z), 
\eeq
where $\phi^{(c)}(z)$ is the normalized chiral $\uu$ boson 
corresponding to the 
charged component of the electron and $\Phi_{\o}(z)$ is a primary field 
($\o$ is a simple current, i.e., $\o* \bar{\o}=1$ \cite{CFT-book,schw}) 
from the neutral sector with CFT dimension $(\o|\o)/2$
\[
\phi^{(c)}(z)\phi^{(c)}(w)\sim -\ln(z-w), \quad 
\Phi_{\o}(z)\Phi_{\o}^*(w)\sim \frac{1}{(z-w)^{(\o|\o)}}.
\]
Next, due to the charge--statistics relation (\ref{0.3b}),  
the conformal  dimension of the electron which is 
\beq\label{dim}
\D_{\mathrm{el}}=\frac{1}{2} (\q_\el|\q_\el)=\frac{1}{2} \left(\frac{1}{\nu_H} + 
(\o|\o) \right) = \frac{1}{2\nu_H}+\D^{(0)}(\o), \quad 
\D^{(0)}(\o)=(\o|\o)\ge 0 \quad
\eeq
must be equal to $p+1/2$ for some $p\in \Z_+$.
According to Eqs.~(\ref{1.3})  and (\ref{dim}) the neutral component 
$\o$ of the electron should  certainly give a nontrivial fractional 
contribution 
to the CFT dimension when the numerator of the filling factor $\nu_H$ is 
bigger than $1$.
Therefore, in general,  the neutral component $\Phi_{\o}(z)$ of the 
electron  field operator cannot be completely removed without 
violating  the fermionic statistics of the electron.
Note that the condition $(\o|\o)\ge 0$ in Eq.~(\ref{dim}) is equivalent to 
$l_{\max} \ge 1/\nu_H$ as formulated in Ref.~\cite{fro-stu-thi}.
\subsection{Spin--charge separation and $\Z_{n_H}$ Pairing rule}
\label{sec:PR}
Because the weight $\o \neq 0$ has at least one non-zero component 
when the numerator of the filling factor is $n_H>1$ the Gram matrix 
(\ref{MSQHL}) in this case is not decomposable, i.e, 
it cannot be represented as a direct sum of two matrices with smaller 
dimensions. This implies that the dual lattice would not be decomposable 
either so that the inequivalent irreducible representations 
of the chiral algebra, i.e., the quasiparticles,  would not be simple tensor 
products of charged and neutral vertex exponents but actually sums of 
such. Therefore the charge and spin are entangled and this absence of 
spin--charge separation makes the situation  more complicated.  
Nevertheless, the spin--charge decomposition 
could be achieved at the expenses of enlarging the 
dual lattice (of excitations) and of introducing a selection rule 
(the $\Z_{n_H}$ {\it pairing rule}).
In fact, there exists a decomposable sublattice 
$L\subset \Gamma$ of the same dimension $N$ spanned by the vectors
\beq\label{Lbasis}
\{ \underline{\e}^1, \a_i \}\ , \qquad 
\underline{\e}^1 = n_H(\q_\el-\o)= \Q \, d_H ,
\eeq
where $\left\{ \a_i \right\}$ are the basis vectors of the neutral 
sublattice $\Gamma_W\subset \Gamma$, i.e., the simple roots 
of the Cartan lattice. This sublattice indeed  splits
into 2 mutually orthogonal sublattices:
\beq\label{1.5}
L=\Z\Q\ d_H \oplus \Gamma_W \ .
\eeq
We have the inclusions $L\subset \Gamma\subset \Gamma^* \subset L^*$
and the physical lattice $\Gamma$ can be decomposed in terms of its 
sublattice $L$ as follows
\beq\label{1.6}
\Gamma=
\left\{ \g=\underline{\l}+s\ \q_\el \; \Big\vert \; \underline{\l}\in L, \; 
0\leq s \leq n_H-1 \right\}, 
\eeq
i.e., $\forall \underline{\gamma}\in\Gamma$,
$\exists !  \, \underline{\lambda}\in L,$ and 
$\exists ! \, s\in\Z_{n_H}$ such that 
$\underline{\gamma}=\underline{\lambda}+s\q_\el$. Note that 
$ L^*/\Gamma^* \simeq \Gamma/L \simeq \Z_{n_H}$ since the determinants of 
the Gram matrices of $L$ and $\Gamma$ (which give the number of sectors of 
the corresponding rational conformal theory) are:
\beq\label{1.7}
|L|=d_H^2 |\Q|^2 |C_{W}|^2=n_H\, d_H \; \det (C_W) =n_H^2 |\Gamma|.
\eeq
The spin--charge separation of excitations is achieved
in the decomposable dual lattice $L^*$, whose \textit{physical points}
(corresponding to the points of $\Gamma^*$) obey a ``$\mod \ n_H$''-selection 
rule:  
if $\underline{\l}\in L^*/L$ then $\underline{\l}$ is a physical 
excitation (i.e., $\underline{\l}\in \Gamma^*/\Gamma$) iff
\[ 
\forall \underline{\l}\in L^*/L :  \quad 
\underline{\l}\in \Gamma^* / \Gamma\subset L^*/L  \quad 
\Longleftrightarrow \quad
\left(\q_\el |\underline{\l}\right)\in \Z.
\]
In other words, if we write the excitation charge $\underline{\l}$ 
in the basis of $L^*$
\beq\label{lambda}
\underline{\l}=l\, \underline{\e}^*_1 +\L, \quad \L=\sum_i a^i\L_i,
\quad l,\, a^i \in \Z
\eeq
with $\L_i$ being the
fundamental weights of the  Witt sublattice, then  $\underline{\l}$ 
is a physical excitation if
\beq\label{Z_n}
\left(\q_\el |\underline{\l}\right)=
\frac{1}{n_H}\left( n_H(\o|\L)-l\right)
 \in \Z \quad 
\Longleftrightarrow \quad  n_H(\o|\L)=l \mod n_H,
\eeq
where $\o=\sum_i \omega^i\L_i$ with $\omega^i \in \Z$
is the neutral component of the electron's topological charge (\ref{1.3}).
This condition reduces the number of elements in $L^*/L$ by a factor of 
$n_H^2$ to that of $\Gamma^*/ \Gamma$ which is the number of  
physically different quasiparticles. To summarize, we first extend 
the dual lattice $\Gamma^*$ to $L^*$ obtaining the spin--charge separation 
in the latter and then identify the physical points in the extended 
lattice $L^*$ by a $\Z_{n_H}$ pairing rule. 
\subsection{The CFT characters for the lattice extensions of $\uu^N$ }
Because of Eq.~(\ref{1.6}) 
the Hilbert spaces $\H^\Gamma$of the original $\Gamma^*/\Gamma$ model 
can be written as direct sums of the Hilbert spaces $\H^L$ of  its 
decomposable sublattice (\ref{1.5})
\[
\H^{\Gamma}_{\underline{\gamma}^*} =
\mathop{\oplus}\limits_{\underline{\gamma}\in\Gamma}
\V_{\underline{\gamma}^*+\underline{\gamma}}=
\mathop{\oplus}\limits_{s\mod n_H} \ \
\mathop{\oplus}\limits_{\underline{\lambda}\in L}
\V_{\underline{\gamma}^*+\underline{\lambda}+s\q_\el}=
\mathop{\oplus}\limits_{s\mod n_H} 
\H^{L}_{\underline{\gamma}^*+\underline{\lambda}+s\q_\el} ,
\]
where $\V_{\underline{\gamma}^*}$ is the $\uu^N$ Verma module
($N\in\N$ is the rank of the lattice $\Gamma$) over the 
lowest-weight state $|\underline{\gamma}^* \rangle$.
Therefore the $\Gamma$-characters would have a similar decomposition
into $L$-characters
\[
\chi_{\underline{\gamma}^*}(\t,\z;\Gamma)=
 \mathop{\mathrm{tr}}_{\ \ \H_{\g^*+s\q_\el}^L \ }
q^{L_0-c/24}\,  \e^{2\pi i \z (\Q|\underline{J}_0)}\ =
\sum_{s\mod n_H} \chi_{\underline{\gamma}^*+s\q_\el}(\t,n_H\z;L),
\]
where the factor $n_H$ in front of $\z$ appeared because $\Q=\q^*_1$
in $\Gamma^*$ but in the decomposable dual lattice it is 
$\Q=n_H\, \underline{\e}^*_1 \in L^*$. 
Because the  sublattice $L$ is decomposable the characters of 
the $\H^L$ modules are simply products of those for the corresponding 
charged and neutral sectors. 
Writing explicitly $\underline{\gamma}^*=l\, \underline{\e}^*_1 +\L$ we have
\beq\label{chi}
\chi_{l,\L}(\t,\z;\Gamma)=\sum_{s\mod n_H} K_{l+sd_H}(\t,n_H\z;n_Hd_H)\
\ch_{\o^s*\L}(\t)
\eeq
where $K_{l}(\t,n_H\z;n_Hd_H)$ is the character of the (charged) 
$\uu$ sector, 
i.e.,  the rational torus partition function \cite{CFT-book} 
\beq\label{K}
K_l(\t,\zeta;m)=\frac{1}{\eta(\t)} \sum_{n\in\Z}
q^{\frac{m}{2}\left(n+\frac{l}{m}\right)^2} 
\e^{2\pi i \zeta \left(n+\frac{l}{m}\right)},
\quad
\eta(\t) = q^{\frac{1}{24}} \prod_{n=1}^\infty \left(1-q^n\right),
\eeq
($\eta$ is the Dedekind function \cite{CFT-book})
while
$\ch_{\L}(\t)$ is that  of the neutral sector of the 
decomposable lattice $L$ 
and $\L$ satisfies  the $\Z_{n_H}$ pairing rule 
$n_H(\o|\L)=l\mod n_H$.
Note that the electron charge vector (\ref{1.3}) acts on the 
neutral sector by the $s$-th power of the  simple current $\o$
which in an abelian CFT is taking the form $\o^s*\L=s\o+\L$.
The characters expressions (\ref{chi}) are valid for the  FQH state 
corresponding to the arbitrary topological charge lattice $\Gamma$. 

\subsection{Neutral reductions and stability criteria}
The classification of the FQH universality classes in terms of  
chiral quantum Hall lattices is motivated by the rationality of the 
filling factors and of the electric charge spectrum. In addition, 
according to Eq.~(\ref{dim}) the neutral component $\o$ of the electron's 
topological charge should also have a rational CFT dimension 
 depending on the filling factor (see Eq.~(\ref{nu-msqhl})). 
That is why the electric properties of the FQH liquid are well-described 
by the topological charge lattice extension of the $\uu^N$ CFT introduced 
in Sect.~\ref{sec:abelian}. 
However the electric properties, which are basically what is measured in 
the experiment, are not enough to fix a unique rational CFT corresponding 
to a given FQH universality class.
For example, we can start with the lattice extension of the $\uu^N$ CFT
and project some degrees of freedom in the neutral sector
keeping  the same electric $\uu$ part. This would certainly preserve the 
electric properties  and produce another candidate CFT for the same 
universality class.
Two important consistency conditions are that the neutral projection 
preserves the neutral component $\o$ of the electron topological charge
 vector, for locality reasons, and the $\Z_{n_H}$ pairing rule which 
is a general property  of the FQH excitations due to the spin--charge 
entanglement.
The advantage of the neutral projections is that  they 
generically reduce the total central charge and therefore may
 produce  better candidates for the same universality class according 
to the stability criteria formulated in Ref.~\cite{fro2000}. 
The neutral projections also preserve the structure 
of the characters (\ref{chi}) only the neutral characters 
$\ch_{\o^s*\L}$ have to be replaced by the corresponding projected 
ones. This is important because the neutral characters give non-trivial
contribution to all thermodynamic quantities.
An example of a neutral projection which we shall describe  
in Sect.~\ref{sec:PF}  is the affine coset reductions which realizes 
the parafermion FQH states in the second Landau level \cite{NPB2001}.

The FQH effect phenomenology seems to be in favor of the neutral 
projections: the main quantum numbers expected  in single-layer disk
QH samples of polarized electrons are electric charge and orbital 
momentum, while in the lattice extensions of  $\uu^N$ there are
$N$ separately conserved charges. The projection would remove the 
superfluous charges keeping only the electric charge and the orbital 
momentum, and  in some cases would  preserve discrete quantum numbers 
related to $W_n$ algebras, like in the coset projection in 
Sect.~\ref{sec:PF}, due to the $\Z_{n_H}$ pairing rule.
\subsection{The Aharonov--Bohm flux: twisting the electron}
\label{sec:AB}
Let us introduce the AB magnetic field piercing the FQH disk sample
at the origin in the $z$ direction which is  produced by a vector 
potential of the form
\beq\label{A_AB}
\mathbf{A}=\frac{h}{e}\frac{\phi}{2\pi r^2}\, (-y,x) = 
\frac{h}{e}\frac{\phi}{2\pi r} \ \mathbf{e_{\varphi}}, \quad 
\mathrm{where} \quad
r=\sqrt{x^2+y^2} 
\eeq
is the length of the radius vector $\mathbf{r}=(x,y)$ in the plane,
$\mathbf{e_{\varphi}}$ is the unit vector along the direction of the polar 
angle $\varphi$  and $\phi\in \R$ is the AB magnetic flux in units of the 
flux quantum $\phi_0=h/e$.
In this section we shall see how the partition function is modified in 
presence of AB flux.

First, because $\mathrm{rot}\,\mathbf{A}=\phi \ \delta^{(2)}(\mathbf{r})$ is 
zero  everywhere 
except for the origin, the electron field $\psi^A_\el(z)$ in presence  
of AB flux can be written as
\beq\label{twist}
\psi^A_\el(z)=\exp\left(-i\frac{e}{\hbar} \int_{*}^{z}
\mathbf{A} . \d\mathbf{r}  \right)
\ \psi_\el(z)=
z^{-\phi}\ \psi_\el(z), \quad \phi \in \R,
\eeq 
($e=\hbar=1$) where $ \psi_\el(z)$ is the electron field operator in the 
absence of 
AB magnetic field, provided that the path of the line integral does not 
encircle the origin. In the computation of the line integral in 
Eq.~(\ref{twist}) we have 
used $z=r\exp(i\varphi)$ and have chosen a path along the $\varphi$ 
coordinate 
line \footnote{in the CFT we conventionally choose $r=1$, however, the 
results can be extended to arbitrary $r$ by analytic continuation of the 
CFT correlation functions}, 
i.e., 
$\d\mathbf{r}=r \, \mathbf{e_{\varphi}}  \d\varphi$, ending 
at $\varphi=\arg(z)$, the polar angle of the coordinate $z$. 
The second equality in Eq.~(\ref{twist}) is a simple special case 
of a more general procedure called twisting \cite{kt,JMP98}.
The AB flux modifies only the $\uu$ part of the electron operator (\ref{el})
which,  for any QH system (integer and fractional), is realized as the  
$\uu$ vertex exponent defined by
\beq\label{vertex}
\np{\e^{ i\alpha \phi^{(c)}(z)}} \ \df
U_\alpha\, \e^{i\alpha \phi^{(c)}_+(z)} \, z^{\alpha J_0} \, 
\e^{i\alpha \phi^{(c)}_-(z)},
\quad 
i\phi^{(c)}_\pm(z)=\pm \sum_{n=1}^\infty J_{\mp n}\frac{z^{\pm n}}{n},
\eeq
where $U_\alpha$ are outer automorphisms of the $\uu$ current algebra 
satisfying  $U_\alpha\, U_\beta=U_{\alpha+\beta}$ for $\alpha,\beta\in\R$, 
$U_0=1$, $\left(U_\alpha\right)^\dagger= U_{-\alpha}$ and 
\beq\label{CR}
\left[J_n,U_\alpha\right]=\alpha \, U_\alpha \, \delta_{n,0}.
\eeq
 Therefore we can use the well known results \cite{kt} for the twisting of 
the normalized $\uu$ current,  $\uu$ Sugawara stress tensor and 
normal ordered exponentials.
The twisting with a twist $\beta$ is  the transformation 
\cite{kt} $\pi_\beta\ : \ A \to  U_{\beta}\, A \, U_{-\beta}$
for any operator $A$ from the chiral (observable) algebra; in particular 
twisting of the vertex exponents gives
\beq\label{e-beta}
\np{\e^{ i\alpha \phi^{(c)}(z)}} \quad 
\stackrel{\pi_\beta}{\longrightarrow} \quad 
U_{\beta} \, \np{\e^{ i\alpha \phi^{(c)}(z)}} \, U_{-\beta} = 
z^{-\alpha\beta}\, 
\np{\e^{ i\alpha \phi^{(c)}(z)}}
\eeq
which can be  computed directly by using the definition ~(\ref{vertex}) 
and the commutation relations~(\ref{CR}) and (\ref{J}).
Because the twisted electron operator 
\beq\label{psi-twist}
\pi_\beta\left(\psi_\el\right)(z)=
\pi_\beta\left(\np{\e^{-\frac{i}{\sqrt{\nu_H}}\phi^{(c)}(z)}} \right)
\Phi_{\underline{\omega}}(z) = 
z^{\frac{1}{\sqrt{\nu_H}}\beta}\psi_\el(z)
\eeq
has to reproduce Eq.~(\ref{twist}) the value of the twist $\beta$ 
corresponding to the addition of AB flux 
$\phi$ can be determined by 
\[
- \left(\frac{-1}{\sqrt{\nu_H}}\right)\, \beta =-\phi\quad \Rightarrow \quad
\beta =-\sqrt{\nu_H}\, \phi.
\]
Then the normalized $\uu$ current (\ref{J}) is modified 
as follows \cite{kt}
\beq\label{J-twist}
J(z) \quad \stackrel{\pi_\beta}{\longrightarrow} \quad 
U_{\beta}\, J(z)\, U_{-\beta} =
J(z)-\frac{\beta}{z},
\eeq
which again can be computed from  the definition
(\ref{J}) of the current and the commutation relations~(\ref{CR}).
 Next, the twist of the electric $\uu$ current (\ref{J_el}) is
\beq\label{Jel-twist}
J^{\mathrm{el}}(z) \quad \stackrel{\pi_\beta}{\longrightarrow} \quad 
J^{\mathrm{el}}(z)+\frac{\nu_H\phi}{z} \quad 
\Longleftrightarrow \quad
J^{\mathrm{el}}_n \quad  \stackrel{\pi_\beta}{\longrightarrow} \quad 
J^{\mathrm{el}}_n+\nu_H\phi \ \delta_{n,0}.
\eeq
Furthermore the twist of the stress tensor $T^{(c)}(z)$ of the $\uu$ part, 
defined by the Sugawara formula (\ref{Tc}), can be derived from 
Eq.~(\ref{J-twist}) to be \cite{kt}
\[
T^{(c)}(z) \quad \stackrel{\pi_\beta}{\longrightarrow}
 \quad  T^{(c)}(z)-\frac{\beta}{z} J(z) +\frac{\beta^2}{2z^2} 
\]
so that the total stress tensor is modified as follows
\beq\label{T-twist}
T(z)\quad  \stackrel{\pi_\beta}{\longrightarrow} \quad 
T(z)+\frac{\phi}{z}J^{\mathrm{el}}(z) +\frac{\nu_H\phi^2}{2z^2}\  
\Longleftrightarrow \
L_n \  \stackrel{\pi_\beta}{\longrightarrow} \ 
L_n+\phi J^{\mathrm{el}}_n +  \nu_H\frac{\phi^2}{2} \delta_{n,0}.
\eeq
\begin{rem} 
The twisted electron $\pi_\beta(\psi_\el)$ has a different CFT 
dimension with respect to the 
untwisted CFT generator $L_0$, i.e.,  different spin and statistics
so that it is not local with respect to $L_0$ and therefore their 
correlation 
functions would not be single-valued in the electron coordinates. 
However, with respect to the twisted generator 
$\pi_\beta(L_0)=L_0+\beta J^\el_0 +\nu_H\phi^2/2$ the twisted electron 
field $\pi_\beta(\psi_\el)$ is local and has the same CFT dimension 
like the untwisted one with respect to $L_0$.  
\end{rem}
Using  Eqs.~(\ref{T-twist}) and (\ref{Jel-twist}) for the zero modes
$J_0^{\mathrm{el}}$ and $L_0$ of the twisted electric current 
and the twisted stress tensor respectively  and putting them into the
 partition function (\ref{Z_chi}) we get the following expression for 
the CFT partition function in presence of AB flux $\phi$
\beqa\label{Z_phi}
Z_{\phi}(\t,\z)&=&
\mathop{\mathrm{tr}}_{\ \ \H \ }
\exp\left\{2\pi i\left[\t\left(L_0 +\phi J_0^\mathrm{el} + 
\nu_H \frac{\phi^2}{2} -\frac{c}{24}\right) +\z \left(J_0^\mathrm{el} + 
\nu_H\phi\right)\right]\right\}
= \nn
&=& \exp\left(2\pi i \nu_H\left[ \frac{\phi^2 }{2}\t +\phi\z\right] \right)\  
Z(\t,\z+\phi\t).
\eeqa
The simple expression (\ref{Z_phi}) for the partition function with 
AB flux  is valid for any quantum Hall state and can be directly used 
for the computation of the free energy (\ref{F}) in presence of flux, 
hence,  of the  persistent currents, AB magnetizations  
and magnetic susceptibilities for temperatures small on the scale of 
the corresponding energy gaps.
Thus Eq.~(\ref{Z_phi}) which is a central result of this paper
is the cornerstone of the effective CFT description of general mesoscopic 
phenomena.
\subsection{Laughlin's spectral flow and Cappelli--Zemba factors}
\label{sec:CZ}
Two of the assumptions in the CFT approach to the FQH effect, based on 
the FQH phenomenology, are that there is a minimal electric charge in 
the quasiparticle spectrum depending on the filling factor and there are 
only a finite number of topologically distinct quasiparticles 
\cite{fro-stu-thi}. This means that the CFT describing the FQH 
states should be rational. While this has not been rigorously proven
it seems reasonable and most of the FQH states for which the CFT are
are well established are indeed rational.
The only exception are the non-rational $W_{1+\infty}$ minimal 
models \cite{ctz} for the Jain series of FQH states. For the 
CFT description of mesoscopic phenomena in FQH systems 
rationality is not necessary,
however  another  condition, called the $V$ invariance
representing the Laughlin spectral flow,  is crucial in this case 
and we shall describe it in this subsection.

The rational CFT characters (\ref{chi_l})
 belong to a finite dimensional representation of the ``fermionic'' 
projective 
 modular group \cite{NPB2001,fro2000,cz} $\Gamma(2)\subset PSL(2,\Z)$, 
i.e., they should be 
$S$ and $T^2$ covariant \cite{cz} where $S$ and $T^2$ are represented by 
the following transformations of the modular parameters
\[
S:(\t,\z)\to \left(-\frac{1}{\t},\frac{\z}{\t}\right), \quad T^2:\t \to \t+2.
\]
In addition, due to the specifics of the  FQH effect there are two more 
covariance conditions to be satisfied by the characters \cite{cz} 
which represent 
certain electric properties of the FQH liquid: the annulus partition function 
must be also invariant under the  $U$  transformation
of the modular parameters  \cite{cz}
\[
U: \  \z\to \z+1
\]
and  under the $V$ transformation
\beq\label{V}
V: \ \z\to\z+\t.
\eeq
The $V$ transformation (\ref{V}) can be interpreted as adding adiabatically
 one flux quantum to the FQH system as can be seen from  Eq.~(\ref{Z_phi})
if we put $\phi=1$. 
Therefore Eq.~(\ref{V}) represents the Laughlin spectral flow 
\cite{laugh,cz}, i.e., the mapping between orbitals, corresponding 
to increasing their orbital momentum  by $1$, when  threading adiabatically
the sample with one quantum of AB flux. This mapping preserves the entire 
quantum spectrum of FQH excitations and should be an exact  symmetry of 
the partition function \cite{laugh,cz}.
In a FQH state, as a result of this flow  a fractional amount of 
electron charge is transferred between the two edges of the  annulus, 
and this is the Laughlin argument describing the  Hall current \cite{laugh}.
However, as noted by Cappelli and Zemba,  the CFT partition function for the 
annulus FQHE sample is not $V$-invariant alone \cite{cz} because  
the transformation~(\ref{V}) changes the absolute values 
of the CFT characters $\chi_\lambda$ in Eq.~(\ref{Z_chi}). 
To preserve the norm of the characters $\chi_\lambda$,  corresponding to a 
general  FQH state, and restore their $V$ covariance,  
 these authors  introduce  special universal non-holomorphic exponential 
factors
\beq\label{CZ}
\exp\left(-{\pi}\nu_H\frac{\left(\Im\z\right)^2}{\Im\t}\right),
\eeq
multiplying the characters (\ref{chi_l}), which we call the 
Cappelli--Zemba (CZ) factors. 
The chiral partition function~(\ref{Z_chi}) also  becomes $V$ invariant 
only after multiplying the characters with the CZ factors (provided 
$\Re\t=\Re\z=0$ as appropriate for the chiral partition function)
\beq\label{Z_CZ}
\widetilde{Z}(\t,\z) \df \e^{-{\pi}\nu_H\frac{\left(\Im\z\right)^2}{\Im\t} }
\, Z(\t,\z) \quad
\Longrightarrow \quad 
\widetilde{Z}(\t,\z+\t)\stackrel{\mathrm{Th}}{=}  \widetilde{Z}(\t,\z).
\eeq
In order to prove the $V$-invariance of   $\widetilde{Z}(\t,\z)$ 
we first use  the following property of the $K$-functions
\beq\label{K-flux}
\e^{-{\pi}\nu_H\frac{\left(\Im(\z+\phi\t)\right)^2}{\Im\t}}
K_l(\t,n_H(\z+\phi\t);n_H d_H)) =
\e^{-{\pi}\nu_H\frac{\left(\Im\z\right)^2}{\Im\t}}
K_{l+n_H\phi}(\t,n_H\z;n_H d_H),
\eeq
 which follows directly from the definition (\ref{K})  
of the $K$ functions and  CZ factors (\ref{CZ}), 
applied for $\phi=1$. Second, because of the $\Z_{n_H}$ pairing rule 
(\ref{Z_n}) it can be shown that the labels of the neutral 
characters $\L= \L_{l \mod n_H}$ are among the fundamental weights
of the Witt sublattice
so  that $\ch_{\o^s*\L}=\ch_{l \mod n_H,s}$ and therefore
$\ch_{l+n_H ,s}=\ch_{l ,s}$ which proves the ``quasiparticle's 
spectral flow''
\beq\label{shift}
\e^{-{\pi}\nu_H\frac{\left(\Im(\z+\phi\t)\right)^2}{\Im\t}}
\chi_{l,\L}(\t,\z+\phi\t)= 
\e^{-{\pi}\nu_H\frac{\left(\Im\z\right)^2}{\Im\t}}
\chi_{l+n_H,\L}(\t,\z).
\eeq
Finally the $V$-invariance of  partition function (\ref{Z_chi})
 follows from the invariance
of the sum under the shift of the $\uu$ label  $l$ (corresponding to
a shift of $\lambda$ in Eq.~(\ref{Z_chi}) ).

The CZ factors were originally interpreted \cite{cz} as adding 
constant capacitive  energy to both edges which makes the ground state 
energy independent of the edge potential $V_0$.
For the single-edge disk sample, however,  it is more intuitive to interpret
the CZ factors directly in terms of the AB flux encircled by the edge.
In this paper we are going to show that these factors play much more 
important role in the equilibrium thermodynamic phenomena, such as the 
persistent currents and AB magnetization. 

Now we can show that the partition function with AB flux (\ref{Z_phi})
is invariant under the Laughlin spectral flow and is essentially  
equivalent to the procedure of Ref.~\cite{cz} of multiplying the 
conventional partition function by the CZ factor. 
First, because the partition functions (\ref{Z_phi}) and 
(\ref{Z_chi}) are chiral (corresponding to a single-edge disk FQH sample) 
the modular parameters $\t$ and $\z$ are purely imaginary, i.e., 
$\Re\t=\Re\z=0$. Then the exponential factor in 
Eq.~(\ref{Z_phi}) can be written equivalently as
\[
\e^{-\pi\nu_H\left[ \phi^2 \,\Im\t+2\phi\,\Im\z\right] }=
\e^{-\pi\nu_H\,\Im\t\left( \phi+ \frac{\Im\z}{\Im\t}\right)^2} \
\e^{\pi\nu_H\,\Im\t\left( \frac{\Im\z}{\Im\t}\right)^2} 
\]
and we note that 
$\exp\left(-\pi\nu_H\,\Im\t\left( \phi+ \frac{\Im\z}{\Im\t}\right)^2\right)=
\exp\left(-\pi\nu_H\,\Im\t\left( \frac{\Im\z'}{\Im\t}\right)^2\right)$ for
$\z'=\z+\phi\t$
so that  the partition function  with AB flux (\ref{Z_phi}) is simply
\beq\label{Z_phi-2}
Z_{\phi}(\t,\z)= \e^{\pi\nu_H\,\Im\t\left( \frac{\Im\z}{\Im\t}\right)^2}\
\widetilde{Z}(\t,\z+\phi\t)
\eeq
where \textit{the exponential prefactor is now $\phi$-independent} and 
$\widetilde{Z}(\t,\z)$ is defined in Eq.~(\ref{Z_CZ}). 
Next, for integer $\phi$, the  invariance of 
$Z_{\phi}(\t,\z)$  under $\phi\to \phi+1$ follows from Eq.~(\ref{Z_CZ}):
\beqa
Z_{\phi+1}(\t,\z)&=&\e^{\pi\nu_H\,\Im\t\left( \frac{\Im\z}{\Im\t}\right)^2}\
\widetilde{Z}(\t,\z+\phi\t+\t) \stackrel{\mathrm{Eq.}~(\ref{Z_CZ})}{=}
\e^{\pi\nu_H\,\Im\t\left( \frac{\Im\z}{\Im\t}\right)^2}\
\widetilde{Z}(\t,\z+\phi\t) = \nn &=&  Z_{\phi}(\t,\z). \nonumber
\eeqa
This result is an illustration  of the  Bloch--Byers--Yang theorem 
\cite{byers-yang} which states that the partition function of any non-simply 
connected conducting system threaded by magnetic flux is a periodic function 
of the flux with period of 1 flux quantum.
As can be seen from Eq.~(\ref{shift}) the spectral flow (\ref{V})
maps one character into another and in order for the  partition function
to be $V$-invariant  it must include all inequivalent sectors which 
are connected by the action of the spectral flow Eq.~(\ref{V}).

To summarize,  the arbitrary flux $\phi$ threading procedure
within the rational CFT framework, is represented by the 
following transformation of the parameters in the partition function
\beq\label{flux}
\z\to\z+\phi \t, \quad \phi\in\R
\eeq
provided that the CZ factors have been included.
Then the partition function for the edge states of a FQH system in 
presence of AB flux represented by the vector potential (\ref{A_AB})
is given by either Eq.~(\ref{Z_phi-2}) or Eq.~(\ref{Z_phi}).
 
Finally we can make the following identification  of the modular parameters
in the context of a disk FQH threaded by AB flux: as we have already seen
$\Re\t=\Re\z=0$, and $\Im\t\sim 1/k_BT$ where the proportionality constant 
sets the temperature scale. We choose the scale $T_0$ from 
Ref.~\cite{geller-loss} so that 
\beq\label{mod_param}
\t=i\pi \frac{T_0}{T},\quad \z=\phi\t, \quad 
T_0=\frac{\hbar v_F}{\pi k_B L}, 
\eeq
where $T$ is the absolute temperature, $L=2\pi R$ is the circumference of 
the edge,  $k_B$ the Boltzmann 
constant and $\phi$ the magnetic flux (in units of $h/e=1$) 
threading the FQH disk.
Note that for $\z=0+\phi\t$, which will be our case, 
 the two partition functions, Eq.~(\ref{Z_phi}) 
and Eq.~(\ref{Z_CZ}) coincide, i.e.,  
$Z_\phi(\t,0)\equiv \widetilde{Z}(\t,\phi\t)$ so that both can be used 
for the 
computation of the persistent currents. 
Let us note  that Eq.~(\ref{flux}) together with
the CZ factors (\ref{CZ}) have already been used in Refs.~\cite{5-2} and 
\cite{PRB-PF_k} for the computation of the persistent currents, AB 
magnetization and susceptibility.
\subsection{Low-temperature asymptotics of the partition function}
\label{sec:Z_low-T}
The low-temperature limit $T\to 0$ in the partition function corresponds 
to the trivial limit $q\to 0$.
Therefore we can keep only the first terms in $q$ in the linear partition
function (\ref{Z_chi})  which come from the vacuum
character and from those with the minimal non-trivial  CFT dimension.
In this paper  we shall assume that the minimal nonzero 
CFT dimensions come from the sectors containing either one quasiparticle (qp) 
or one quasihole (qh) but not both. In this case the partition function 
can be approximated by
\beqa\label{Z_chi_0}
\widetilde{Z}(\t,\z) &\mathop{=}_{\ T\ll T_0 \ } &
	\e^{-{\pi}\nu_H\frac{\left(\Im\z\right)^2}{\Im\t}}
\Bigl( K_{0}(\t,n_H\z;n_H d_H)\, \ch_0(\t)+ \nn
& + & K_{1}(\t,n_H\z;n_H d_H)\, \ch_\qh(\t) +  
K_{-1}(\t,n_H\z;n_H d_H)\,\ch_\qp(\t)\Bigr),
\eeqa
where $n_H$ and $d_H$ are the numerator and denominator of the filling factor
(\ref{0.3}), respectively.
This  assumption is natural for the FQH states for which the 
CFT dimension of the quasihole/quasiparticle, which carry the minimal 
absolute value of the electric charge (corresponding to $l=\pm 1$ in the $K$ 
functions) in the spectrum,  is the minimal one allowed in the CFT.
Notice that the maximally symmetric quantum  Hall lattice  CFT
for the Jain series labelled by $(2p+1|{}^1A_{n-1})$ does not satisfy 
this assumption,
however this is not a problem since it is almost certain that 
this CFT is not the correct one \cite{grayson}.
In Eq.~(\ref{Z_chi_0}) $\ch_0(\t)$ is the CFT character of the vacuum 
representation of the neutral sector while $\ch_\qh(\t)$ and  $\ch_\qp(\t)$
are those of the neutral sectors containing one quasihole and one 
quasiparticle, respectively and we note that only the leading 
powers of $q$ in the characters are important for $T\to 0$.

To include the AB flux we apply the arbitrary flux transformation 
(\ref{flux}) to the partition function (\ref{Z_chi_0}) and  use the 
property (\ref{K-flux}) of the $K$-functions (\ref{K}) for $\z=0$. 
Moving the flux dependence into the indices of the $K$-functions 
is technically more convenient since it guarantees the reality of the 
partition functions when computed numerically for arbitrary $\phi$.
 Keeping only the first terms in Eq.~(\ref{Z_chi_0})
and skipping the $\phi$-independent  terms \footnote{for the computation 
of the  the persistent currents  we intend to differentiate the logarithm of 
the partition function  with respect to $\phi$ according to 
Eqs.~(\ref{F}) and (\ref{I}) so that the $\phi$-independent 
terms would not contribute} containing the
$\eta$-function and the central charge we can write\footnote{after
including the flux the indices of the $K$ functions are modified so
that e.g., $K_{2+n_H\phi}$ may dominate over $K_{-1+n_H\phi}$ for
$-1/2 < \phi<0$ in the $T\to 0$ limit. However, in this case both
functions can be neglected when compared to $K_{1+n_H\phi}$ since
the latter gives the smallest minimal CFT dimension }
\beqa\label{Z_0}
\widetilde{Z}(T,\phi)  &\mathop{=}_{\ T\ll T_0 \ }&
q^{\frac{\nu_H}{2}\phi^2} \left\{ 1+ q^{\D_\qh}
\left( q^{Q_\qh\phi}+q^{-Q_\qh\phi}\right)\right\}  = \nn
&=&  \e^{- \pi^2 \frac{T_0}{T}\nu_H\phi^2}
\left\{ 1+ 2 \e^{-\frac{\widetilde{\eps}_\qh}{k_B T}}
\cosh\left(2 \pi^2 \frac{T_0}{T}Q_\qh\phi\right)    \right\}
\eeqa
where
\beq\label{proper}
\widetilde{\eps}_{\qh}= 2 \pi^2 k_B T_0 \D_\qh, \quad
 \D_\qh=\frac{1}{2n_H d_H} +\D^{(0)}_\qh ,\quad
 Q_\qh=\frac{1}{d_H}.
 \eeq
In the above equation  $\widetilde{\eps}_{\qh}$ is the
proper quasihole energy,   $\D_\qh$ is the total
CFT dimension of the quasihole (with  $\D^{(0)}_\qh$  being its
neutral component)  and $Q_\qh$ is the quasiholes electric charge.
In the derivation of Eq.~(\ref{Z_0}) we have also used
$\ch_0(\t)\simeq q^0$, $\ch_\qp(\t)=\ch_\qh(\t)\simeq q^{\D^{(0)}_\qh}$
for $q\to 0$ (again dropping the central charge term).
\begin{rem}
The partition function (\ref{Z_0})
is finite in this approximation since for all
$\phi$ one has
$\nu_H \phi^2/2  \pm Q_\qh \phi +\D_\qh \geq 0$
because the discriminant of the above quadratic
equation is $-2\nu_H \D^{(0)}_\qh\leq 0$ provided that
$\D^{(0)}_\qh\geq 0$.
However, when in the low-temperature limit 
we take the logarithm in Eq.~(\ref{F}) we are 
going to use the
approximate expansion $\ln(1+X)\simeq X$ which is valid for $X\ll 1$.
In our case, for $T\ll T_0$
\[
X=  2 \, \e^{-\frac{\widetilde{\eps}_\qh}{k_B T}}
\cosh\left(-2 \pi^2 \frac{T_0}{T}Q_\qh\phi\right)   \ll 1
\quad \mathrm{iff} \  \left|\phi\right| <  \frac{1}{2}
\ \mathrm{and} \ \D_\qh -\frac{1}{2}Q_\qh \geq 0.
\]
The last condition is always fulfilled in the FQH states,
such as the Laughlin states and the $\Z_k$ parafermion states 
which satisfy the quasihole charge--statistics relation \cite{gaps}
\beq\label{c-s}
2\D_\qh=Q_\qh.
\eeq
\end{rem}
As we shall see in Sect.~\ref{sec:low-T} the low-temperature
asymptotics of the persistent current's amplitude  derived from 
Eq.~(\ref{Z_0}) is logarithmic
\beq\label{I_max2}
I_{\max}(T)\  \mathop{=}_{T\ll T_0} \ \frac{1}{2}\nu_H   \frac{\e v_F}{L}
\left\{ 1 -\frac{k_B T}{\widetilde{\eps}_\qh}
	\left[1+\ln\left( \frac{2}{n_H}\frac{\widetilde{\eps}_\qh}{k_B T}\right) \right]
	\right\}
\eeq
and shows that the persistent current decays logarithmically
with increasing temperature due to the probability for thermal activation
of quasiparticle--quasihole pairs\footnote{note that the proper quasihole
energy appears with a factor of $2$} which reduces the radial electric
field that is responsible for the appearance of the azimuthal persistent
current.
\subsection{High-temperature asymptotics: modular $S$ matrix and 
quantum dimensions of quasiparticles}\label{sec:Z_high-T}
The high-temperature limit $T\to \infty$ is rather non-trivial
since the modular parameter $q=\exp(-2\pi^2 T_0/T) \to 1$ is at the 
border of the
convergence interval for the partition functions and therefore cannot be
taken directly. However, one could use the advantage that the
complete chiral partition function is constructed as a sum of RCFT characters,
so that
\[
\chi_{\l}(\t,\z)= \sum_{\l'=1}^N S_{\l\l'} \
\chi_{\l'}(\t',\z'),
\]
where the modular parameters $(\t,\z)$ and $(\t',\z')$ are related by
\beq\label{new-param}
S:
\left|
\begin{array}{l}
\t=-\frac{1}{\t'} \\
\z=-\frac{\z'}{\t'} \end{array} \right.
\quad
{\Longleftrightarrow} \quad
\left|
\begin{array}{l}
\t'=-\frac{1 }{\t} =i \frac{T}{\pi T_0} \\
\z'=\frac{\z}{\t} = \phi \end{array}\right. .
\eeq
Apparently Eq.~(\ref{new-param}) connects the 
high-temperature  and low-temperature limits  of the partition 
function. 
Note that the usual common phase factor \cite{cz} in front of the
$S$ matrix
is trivial since $\exp\left(i\pi\nu_H \Re({\z'}^2/{\t'})\right)=1$.
Now the new modular parameter vanishes in the high-temperature limit
\beq\label{q'}
q'=\e^{2\pi i \t'}= \exp\left(-\frac{2T}{T_0} \right) \to 0 \quad
\mathrm{when} \quad T\to \infty
\eeq
so that it would be enough to keep only the leading terms.
After the $S$ transformation~(\ref{new-param}) the partition 
function (\ref{Z_chi}) 
becomes \footnote{the CZ factors (\ref{CZ}) are trivial 
after $S$ transformation 
$\exp( - \pi\nu_H(\Im\z')^2/\Im\t')=1$ since $\z'=\phi$, i.e., 
$\Im\z'=0$ }
\beq\label{Z-high}
Z(\t,\z)= \sum_{\l=1}^N F_\l\, \chi_{\l}(\t',\z') , \quad
 F_\l= \sum_{\l'=1}^N S_{\l\l'}.
\eeq
Obviously the universal quantities $F_\l$ are supposed to play an
 important role in the high-temperature analysis. 
\begin{rem}
Note that $F_\l$ are always real. The reason is that 
 the $S$ matrix is symmetric 
$S_{\lambda,\rho}= S_{\rho,\lambda}$ and
satisfies $S_{\lambda,\rho^*}= \left(S_{\rho,\lambda}\right)^*$ 
\cite{CFT-book} where 
$\rho^*$ is the the label of the  charge conjugated representation 
of $|\rho\rangle$, i.e., 
$|\rho^*\rangle= C |\rho\rangle$ with $C=S^2$. 
Therefore, for self-conjugated weights, $\rho^*=\rho$, the $S$ matrix has 
real rows and columns $S_{\lambda,\rho}= \left(S_{\lambda,\rho}\right)^*$
(for all $\lambda$)
while for $\rho^*=\sigma\neq \rho$ the rows and columns corresponding to 
$\rho$ and $\sigma$ are complex conjugated
$S_{\lambda,\sigma}= \left(S_{\lambda,\rho}\right)^*$ (for all $\lambda$).
Because the representations 
are either self-conjugate or enter in  conjugated pairs 
the sum of the elements of an entire row or column, i.e., over 
 over all representations, will always be real.
\end{rem}
Since the complete CFT contains a $\uu$
 factor it follows from the properties of the simple currents that 
\[
F_\l \neq 0 \quad \Longleftrightarrow \quad Q_{\mathrm{el}}(\l)=0.
\]
 In particular $F_1$, where $1$ denotes the vacuum representation,
 is the following sum 
 \[
 F_1=S_{11}\sum_{\lambda=1}^N D_\lambda, \quad \mathrm{where}\quad 
D_\lambda=\frac{S_{1\lambda}}{S_{11}}>0
\] 
are the quantum
 dimensions of the topological excitations of the FQH fluid.
Note that the quantum dimensions have several nice properties 
which justify the name ``dimensions''
\[
D_{\mu}D_{\nu}=\sum_\rho N_{\mu\nu}{}^\rho D_{\rho}, \quad
\sum_\lambda D_{\lambda}^2=\left(S_{11}\right)^{-2} 
\]
where $N_{\mu\nu}{}^\rho$ are the fusion coefficients \cite{CFT-book}.
The second of the above equalities, which follows from the unitarity 
of the $S$ matrix,  shows that the vacuum in a rational 
CFT does know about all topologically distinct representations because
the matrix element $S_{11}$ depends only on the vacuum module.

Using Eq.~(\ref{Z-high}) we are going to show in Sect.~\ref{sec:high-T} that 
the high-temperature asymptotics of the persistent current's amplitude 
in general has the form 
 \beq\label{high-T-gen}
I^{\max}(T) \ \mathop{\simeq}_{T\gg T_0} \ I_0 \, 
\left(\frac{T}{T_0}\right) \, \exp\left(-\alpha\frac{T}{T_0}\right),
 \eeq
with a   universal exponent $\alpha$ which can be written in 
the form
\[
 \alpha = \frac{1}{\nu_H}+2\D^{(0)}\left(\L_*\right),
\]
where $\L_*$ is the neutral weight  with the minimal CFT dimension  
satisfying the pairing rule $P[\L_*]\equiv d_H \mod n_H$, see 
Eq.~(\ref{alpha}). According to Eq.~(\ref{high-T-gen}) 
at high temperatures  the mechanism suppressing the 
persistent currents is different from the low-temperature one: 
the amplitudes of these currents decay 
exponentially due to the effects of thermal smearing and decoherence.
\section{Application:  the  $\Z_k$ parafermion FQH states}
\label{sec:PF}
\subsection{The CFT partition function for the  $\Z_k$ parafermion 
FQH states}
Following Refs.~\cite{NPB2001,gaps} we denote the CFT for the 
$\Z_k$ parafermion FQH states ($\PF_k$) in the second Landau level by 
\cite{NPB2001,gaps}
\beq\label{u1xPF}
\left(\widehat{u(1)}\oplus \PF_k \right)^{\Z_k}, \quad 
\nu_H=\frac{k}{k+2}
\eeq
where  $\nu_H$ is the filling factor in the last occupied Landau level.
This CFT contains the standard $\uu$ factor describing the charge/flux 
degrees 
of freedom and a neutral $\Z_k$ parafermion sector which is realized as a 
diagonal affine coset \cite{NPB2001}
\beq\label{PF_k}
\PF_k = \frac{\widehat{su(k)}_1\oplus\widehat{su(k)}_1}{ 
\widehat{su(k)}_2}, \quad
c_\PF^{(0)}=\frac{2(k-1)}{k+2}
\eeq
where $c_\PF^{(0)}$ is the central charge of the coset.
The affine coset realization of the parafermions has a long history,
please see Refs.~\cite{halpern-PF,halpern,cristofano-PF} and the references in
\cite{NPB2001}.
The primary fields of the coset (\ref{PF_k}) are labelled by \cite{NPB2001}
\beq\label{PF_labels}
\L_\mu+\L_\rho, \quad 0\leq\mu\leq\rho\leq k-1,
\eeq
where $\L_\mu$, for $\mu=0, \ldots, k-1$ are the $su(k)$ fundamental 
weights ($\L_0=0$).
Because the numerator of the filling factor is $k$  the two factors 
in Eq.~(\ref{u1xPF}) are subjected to a $\Z_k$ pairing rule \cite{NPB2001}, 
in agreement with the general  discussion in Sect.~\ref{sec:PR},
i.e.,  an excitation
\beq\label{primary}
\Phi(l;\L_\mu+\L_\rho)=
\np{\e^{i\frac{l}{\sqrt{k(k+2)}}\phi^{(c)}(z)}} \otimes \ 
\Phi^\PF(\L_\mu+\L_\rho)(z), 
\quad 0\leq\mu\leq\rho\leq k-1
\eeq 
of the $\PF_k$ model with labels $(l,\L_\mu+\L_\rho)$, where $l$ is the 
$\uu$ label and $\Phi^\PF\left(\L_\mu+\L_\rho\right)$ is a 
$\Z_k$-parafermion (primary) field 
\cite{NPB2001,CFT-book}, is legal only  if
\[
P\left[ \Phi^\PF\left(\L_\mu+\L_\rho\right)\right]\equiv l \ \mod \ k, 
\quad\mathrm{where}\quad
P\left[ \Phi^\PF\left(\L_\mu+\L_\rho\right)\right]\equiv \mu+\rho \mod k
\]
is the parafermion $\Z_k$ charge.
The CFT dimensions of the coset primary fields  
$\Phi^\PF\left(\L_\mu+\L_\rho\right)$ have been found in Ref.~\cite{NPB2001}
in a compact form
\beq\label{D_PF}
\D^\PF\left(\L_\mu+\L_\rho\right)=\frac{\mu(k-\rho)}{k}+
\frac{(\rho-\mu)\left[k-(\rho-\mu)\right]}{2k(k+2)},\quad
0\leq\mu\leq\rho\leq k-1.
\eeq
The independent characters $\chi_{l,\rho}$ of the chiral CFT
for the 
parafermion FQH states  
are labelled \cite{NPB2001} by a pair of integer numbers  $(l,\rho)$ 
restricted by
\beq\label{l-rho}
l \mod k+2\quad \mathrm{and} \quad \rho=0,\ldots,k-1,
\quad \mathrm{where} \quad
\rho\geq l-\rho \mod k
\eeq
which determines the topological order to be $(k+1)(k+2)/2$. 
The chiral partition function with the CZ factors~(\ref{CZ}) 
sitting  explicitly  in front of it is
\beq\label{Z_PF}
\widetilde{Z}_{k}(\t,\z) =
\e^{-{\pi}\nu_H\frac{\left(\Im\z\right)^2}{\Im\t}}
\sum_{l  \mod  k+2} \quad
\sum_{\rho \geq l-\rho \mod k } \chi_{l,\rho}(\t,\z),
\eeq
where
\beq\label{chi_PF}
\chi_{l,\rho}(\t,\z) =
\sum_{s=0}^{k-1} K_{l+s(k+2)}(\t,k\z;k(k+2))\,
\ch\left(\L_{l-\rho+s} +\L_{\rho+s}\right)(\t)
\eeq
The characters $K_{l}(\t,k\z;k(k+2))$ of the $\uu$ sector
are defined by Eq.~(\ref{K}).
The  neutral characters 
$\ch(\L_\mu+\L_\rho)$ which are the characters of the coset (\ref{PF_k})
can be found using the general approach \cite{halpern}
are labeled by the weights (\ref{PF_labels}) and following 
Ref.~\cite{NPB2001} can be expressed
in terms of the so called universal chiral partition functions
(see Ref.~\cite{NPB2001} and the references therein) $\ch^Q_\s$ 
after the identification
$Q=\rho$, $\s = \rho-\mu$
\beq\label{ch}
\ch^Q_{\s}(\t)=q^{\D_\s-\frac{c_{\PF}}{24}}
\sum_{
\mathop{n_1,n_2,\ldots,n_{k-1}=0}\limits_{
	 [\underline{n}]	\equiv Q \mod k}}^\infty
	\frac{q^{\underline{n}\cdot C^{-1}\cdot (\underline{n}- \L_{\s} ) }}
{ (q)_{n_1}\cdots (q)_{n_{k-1}}},
\quad 0\leq\s\leq Q.
\eeq
In the above equation, $\underline{n} =(n_1,n_2,\ldots,n_{k-1})$ is
a $k-1$ component vector with non-negative entries,
 $[\underline{n}]=\sum_{i=0}^{k-1} i \, n_i$ is its $k$-ality,
 $\D_\s$ is the  CFT dimension (\ref{D_PF}) of the
 $\PF_k$ primary field labeled by\cite{NPB2001} $\L_0+\L_\s$,
$c_{\PF}=c_{\PF}^{(0)}$ is the parafermion central charge from 
Eq.~(\ref{PF_k}),
\[
 (q)_n=\prod\limits_{j=1}^n (1-q^j)
\]
and $C^{-1}$ is the inverse of the $su(k)$ Cartan matrix.
The partition functions with AB flux, 
is computed according to Sect.~\ref{sec:AB} by applying the 
flux transformation (\ref{flux}) to the partition 
function (\ref{Z_PF})and  using  again the property 
 (\ref{K-flux}) of the $K$-functions for $\z=0$. 
In Sect.~\ref{sec:pers-CFT} we shall use the CFT partition 
function~(\ref{Z_PF}) with AB flux  to compute numerically 
the persistent currents in the parafermion FQH states 
for a wide range of temperatures as well as to 
find analytical formulas for the low- and high- temperature limits. 
\subsection{The modular $S$ matrix for the $\Z_k$-parafermion FQH 
states}\label{sec:S}
For the analysis of the high-temperature asymptotics of the persistent 
currents in the parafermion FQH states we shall explicitly need the 
modular $S$ matrix for the corresponding CFT.
The computation of the $S$ matrix for the $\widehat{u(1)}\times \PF_k$ 
model is a bit  complicated by the fact that the characters (\ref{chi_PF})
are not simply products of characters of its neutral and charged sectors 
 but  sums of these. Nevertheless the $S$ matrix can be found by carefully 
using the properties of the simple currents  \cite{CFT-book,schw}
and following the standard Goddard--Kent--Olive approach 
\cite{gko,CFT-book} for the coset.
In order to compute the full $S$ matrix  we note 
that the full $\PF_k$ characters (\ref{chi_PF}) can be written 
in terms of the simple current 
\beq\label{simp}
\J=\ \np{\exp\left( i \frac{k+2}{\sqrt{k(k+2)}} \phi^{(c)}(z)\right)}
\otimes \ \Phi^{\PF}(\L_1+\L_1)(z),
\eeq
(coinciding with the physical hole operator) which  acts on the primary 
fields (\ref{primary}) according to
\[
\J* \left(\np{\e^{i\frac{\l}{\sqrt{k(k+2)}}\phi^{(c)}(z)}} \otimes \ 
\Phi^\PF(\L_\mu+\L_\rho)(z) \right) =
\np{\e^{i\frac{\l+(k+2)}{\sqrt{k(k+2)}}\phi^{(c)}(z)}} \otimes \ 
\Phi^\PF(\L_{\mu+1}+\L_{\rho+1})(z),
\]
as the sum over the orbit of the product 
$K_l \ \ch(\L_{l-\rho}+\L_\rho)$, i.e., 
\[
\chi_{l,\rho}(\t,\z)= \sum_{s=0}^{k-1} \J^s *
\left(  K_l\left(\t,k\z;k(k+2)\right)
\ \ch(\L_{l-\rho}+\L_\rho) (\t) \right) .
\]
Next we use the general property  of the simple currents' action over 
the $S$ matrices \cite{CFT-book,schw}:
\beq\label{simp_prop}
S_{\J*\Lambda,M}= \e^{-2\pi i Q_J(M)}S_{\Lambda,M}, \quad
Q_{\J}(M) = \D(\J*M)-\D(M)-\D(\J) \mod \Z
\eeq
where $\Lambda$ and $M$ are the  labels of any two coset primary fields,
 $\D(\Lambda)$ is the total CFT dimension of the field labeled by 
$\Lambda$ and $Q_{\J}(M)$ is the monodromy charge \cite{schw}.
Because $\J$ is a simple current the fusion product $\J*M$ contains 
only one primary field so that the monodromy charge is well defined.
Then we apply the
modular  $S$ transformation to a single product of characters 
$K_l \ \ch(\L_{l-\rho}+\L_\rho)$, namely the first term $s=0$ in  
the character (\ref{chi_PF}) which gives
\beqa
S K_l(\t,k\z;k(k+2)) \ \ch(\L_{l-\rho}+\L_\rho)(\t) &=&
\e^{i\pi\nu_H \Re\left( \frac{\z^2}{\t}\right)}
\sum_{l'\mod k(k+2)} S_{l,l'}^{(1)} K_{l'}(\t,k\z;k(k+2))  \nn &\times&
\sum_{0\leq\mu\leq\nu\leq k-1} \Sc^{l-\rho,\rho}_{\mu,\nu} 
\ch(\L_{\mu}+\L_\nu)(\t), \nonumber
\eeqa
where 
\beq\label{S1}
S_{l,l'}^{(1)}=\frac{1}{\sqrt{k(k+2)}}
\exp\left(-2\pi i \frac{l l'}{k(k+2)}\right),
\quad\mathrm{with}\quad  l,l' \mod k(k+2),
\eeq
is the $S$ matrix for the $\widehat{u(1)}_{k(k+2)}$ and 
$\Sc^{\rho,\s}_{\mu,\nu}$ is the one for the diagonal coset 
(\ref{PF_k}) which will be discussed in more detail in 
Appendix~\ref{app:coset_S}. Next we apply the $S$ 
transformation to the full character (\ref{chi_PF}) using the 
property (\ref{simp_prop}) of the simple currents (\ref{simp})
to get
\beqa
S\chi_{l,\rho}&=&\e^{i\pi\nu_H \Re\left( \frac{\z^2}{\t}\right)}
\sum_{s=0}^{k-1} \sum_{l'\mod k(k+2)}\sum_{0\leq\mu\leq\nu\leq k-1}
\exp\left[-2\pi i\left(Q^{(1)}_{\J^s}(l')+\Qc_{\J^s}(\L_\mu+\L_\nu) \right)
\right]
 \times \nn
&\times& S^{(1)}_{ll'} \ \Sc^{l-\rho,\rho}_{\mu,\nu}
 K_{l'}(\t,k\z;k(k+2)) \ \ch\left(\L_\mu+\L_\nu\right),
\label{Sx}
\eeqa
where the monodromy charge $Q^{(1)}$ for the $u(1)$ part is
(using $\D^{(1)}(l)=l^2/2k(k+2)$ )
\[
Q^{(1)}_{\J^s}(l')=\D^{(1)}\left(l'+s(k+2)\right) -
\D^{(1)}\left(l'\right) - \D^{(1)}\left(s(k+2)\right)
 =\frac{sl'}{k},
\]
and that for the coset part is (using Eq.~(\ref{D_PF}) for 
the CFT dimension in the coset)
\[
\Qc_{\J^s}=\D^\PF\left(\L_{\mu+s}+ \L_{\nu+s}\right)  -
\D^\PF\left(\L_{\mu}+ \L_{\nu}\right) -
\D^\PF\left(\L_{s}+ \L_{s}\right)
=-\frac{s(\mu+\nu)}{k}.
\]
Because the only $s$-dependence is under the exponent of the 
monodromy charges the sum over $s$ can be taken explicitly and 
  gives a $\delta$-function
\[
\sum_{s=0}^{k-1} \exp\left(-2\pi i \frac{s(l'-\mu-\nu)}{k}\right)=
k \  \delta\left(l'=\mu+\nu \mod k\right).
\]
Plugging this $\delta$-function back in Eq.~(\ref{Sx}) and splitting 
the summation variable $l' \mod k(k+2)$ into
\[
l'=l'' +s''(k+2), \quad \mathrm{where} \quad s'' \mod k
\quad \mathrm{and} \quad l'' \mod k+2,
\]
we get (setting  $\nu=:\rho'$ $\Rightarrow$ $\mu=l'-\rho' \mod k \leq \rho'$)
\beqa
S\chi_{l,\rho}&=&\e^{i\pi\nu_H \Re\left( \frac{\z^2}{\t}\right)}
\sum_{l''\mod k+2}\sum_{s=0}^{k-1} \sum_{\rho'=0}^{k-1}
 S^{(1)}_{l,l''+s''(k+2)} \ 
\Sc^{l-\rho,\rho}_{l''-\rho'+s''(k+2),\rho'} \times \nn
&\times& K_{l''+s''(k+2)}(\t,k\z;k(k+2)) \ 
\ch\left(\L_{l''+s''(k+2)-\rho'}+\L_{\rho'}\right).
\nonumber
\eeqa
In the above equation we can change $s''-\rho'=:\rho''$ and interpret 
the sum over $s''$ as a simple current's orbit because
\[
\L_{l''+s''(k+2)-\rho'}+\L_{\rho'}=
\L_{l''-\rho'' +s''}+\L_{\rho''+s''}=J^{s''}*
\left(\L_{l''-\rho''}+\L_{\rho''}\right).
\]
Using once again the property (\ref{simp_prop}) of the $S$ matrix 
under the action of the simple currents, i.e.,
\[
S^{(1)}_{l,l''+s''(k+2)}=
\exp\left(-2\pi i \frac{ls''}{k}\right) S^{(1)}_{l,l''}
\quad\mathrm{and}\quad
\Sc^{l-\rho,\rho}_{l''-\rho''+s'',\rho''}=
\exp\left(-2\pi i \frac{-ls''}{k}\right)
\Sc^{l-\rho,\rho}_{l''-\rho'',\rho''} 
\]
one could see that the two exponential factors cancel each other
and the sum over $s''$ exactly reproduces the character 
$\chi_{l'',\rho''}$ according to Eq.~(\ref{chi_PF}). 
Thus we finally  get the  following compact  expression 
for the total $S$ matrix of the $\uu \times \PF_k$ model
\beq\label{S_PF}
S\chi_{l,\rho}={\sum_{l''\mod k+2}}' \ \sum_{\rho''=0}^{k-1}
S^{l,\rho}_{l'',\rho''}\ \chi_{l'',\rho''}  \quad \mathrm{with} \quad
S^{l,\rho}_{l'',\rho''} =k \ S^{(1)}_{l,l''} \ 
\Sc^{l-\rho,\rho}_{l''-\rho'',\rho''}, 
\eeq
where  $l-\rho \mod k \leq \rho$ and the prime over the first sum
 means that the summation over $l''$ and $\rho''$ 
is restricted by $(l''-\rho'' \mod k) \leq \rho''$;
the $\uu$ modular matrix $S^{(1)}$ is given in Eq.~(\ref{S1}) 
while the coset $S$ matrix can be written explicitly in the basis of
coset primary fields $\Phi^\PF(\L_\mu+\L_\nu)$  labelled by 
$\L_\mu+\L_\nu$ with $0\leq\mu\leq\nu\leq k-1$ as 
\beq\label{Sc}
\Sc^{\L_\mu + \L_\nu}_{\L_\rho + \L_\sigma} =
\Sc^{\mu,\nu}_{\rho,\sigma}=
\e^{2\pi i\frac{(\mu+\nu)(\rho+\sigma)}{k}}
\ \left(  S^{(2)}_{\L_\mu + \L_\nu, \L_\rho + \L_\sigma}\right)^{*},
\quad \left\{ 
\begin{array}{ll}  0\leq\mu\leq\nu\leq k-1 \\ 
0\leq\rho\leq\sigma\leq k-1\end{array}\right. .
\eeq
In Eq.~(\ref{Sc})  $S^{(2)}_{\L,\L'} $ is the $S$ matrix for 
$\widehat{su(k)}_2$ which can be computed as a sum over the 
$su(k)$ Weyl group \cite{kac}
\beq\label{Kac}
S^{(2)}_{\L;\ \L'}=\frac{i^{\, k(k-1)/2}}{\sqrt{k(k+2)^{k-1}}}
\sum_{w\in \mathcal{W}} \epsilon(w) \
\exp\left(  - 2\pi i
\frac{\left(\L+\underline{\rho}
\left|w(\L'+\underline{\rho})\right)\right.}{k+2}\right)
\eeq
The derivation of diagonal coset $S$ matrix (\ref{Sc}) following the 
Goddard--Kent--Olive construction \cite{gko,CFT-book} is sketched in 
Appendix~\ref{app:coset_S}.
Eq.~(\ref{Sc}) shows that the $S$ matrix for the diagonal coset
(\ref{PF_k}) is related to that of the $\widehat{su(k)}_2$ current algebra,
by charge conjugation and the action  of simple currents 
which are automorphisms of the fusion rules, i.e., 
the two $S$ matrices are equivalent.  Therefore not only the 
irreducible representations of the coset (\ref{PF_k}) and 
$\widehat{su(k)}_2$ are labelled by the same weights
(\ref{PF_labels})  but the fusion rules of both CFTs are identical.
This explains why the quantum group structure of the affine coset
(\ref{PF_k})  
is the same as that for $\widehat{su(k)_2}$
 (see Ref.~\cite{diag_coset} and the references therein), i.e., 
\[
U_q\left( sl(k)\right) \quad \mathrm{with} \quad
q=\exp\left(-i\frac{\pi}{k+2}\right),  \quad
{\lfloor m\rfloor}_q=\frac{q^{m}-q^{-m}}{q^{1}-q^{-1}}.
\]
\section{Persistent currents}
\subsection{An intuitive picture}
\begin{figure}[htb]
\begin{center}
\epsfig{file=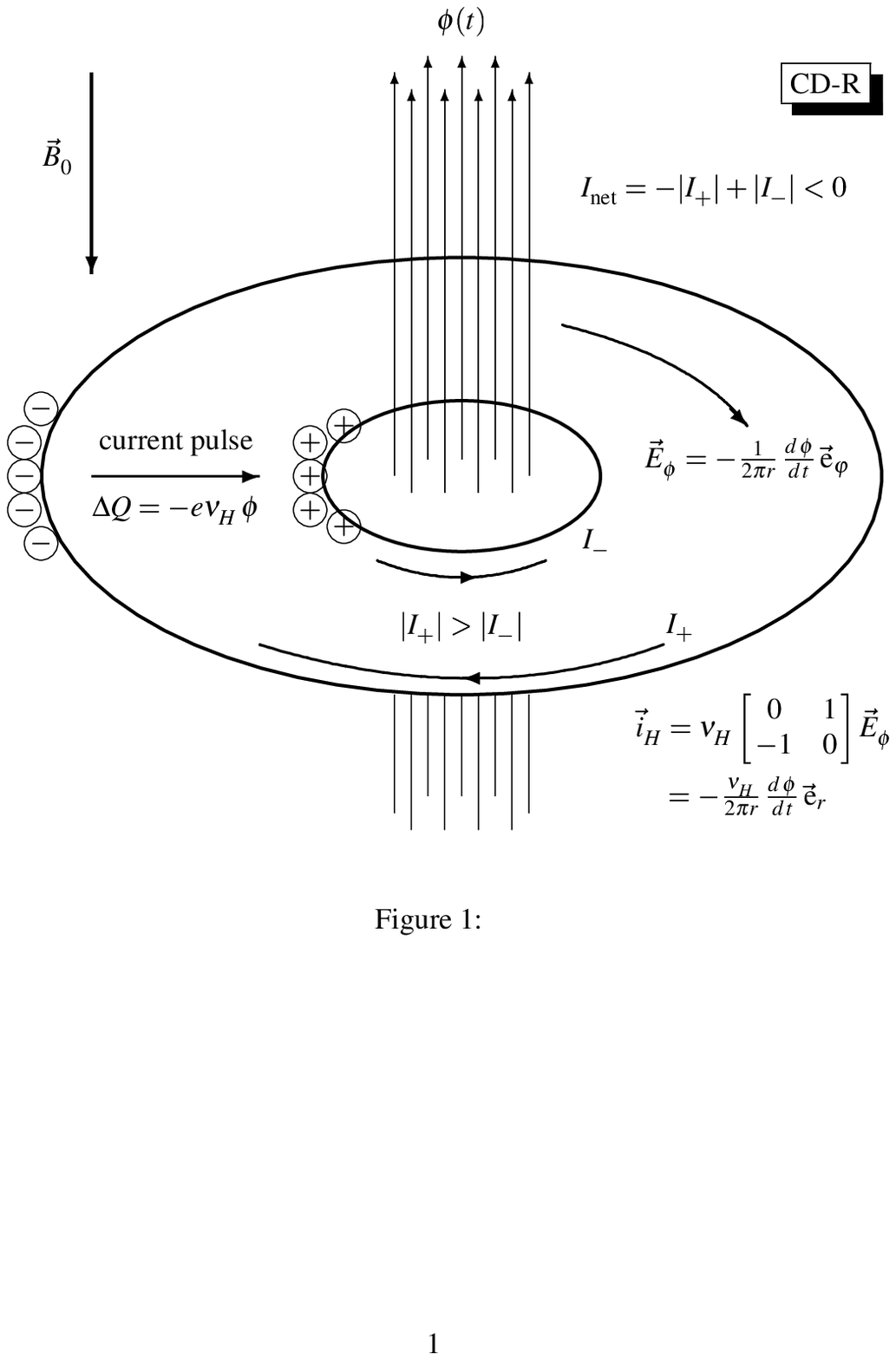,height=9cm,clip=,%
bbllx=150,bblly=300,bburx=470,bbury=600}
\caption{The two-dimensional Corbino disk in ring geometry (CD-R) 
and the persistent currents on both edges \label{fig:corbino}}
\end{center}
\end{figure}

Consider  the Corbino ring sample shown in Fig.~\ref{fig:corbino}  
threaded by AB flux that is changed adiabatically at zero temperature
\footnote{this process is  reversible on each step}.
We shall assume that the huge  background homogeneous  magnetic field 
\mbox{$\mathbf{B}_0=-B_0\,{\bf e}_z$} (perpendicular to the plane), 
which points in the negative $z$ 
direction, has driven 
the electron system into an incompressible FQH state on the ring
so that $\sigma_{xx}=0$ and $\sigma_{xy}=\nu_H e^2/h$. 
The actual magnetic field is measured with respect to $B_0$, i.e., 
$B=B_{\mathrm{tot}}- B_0$.
When no flux is threading the sample, i.e., $B=0$, the currents are 
concentrated on the edges and have the same magnitude but opposite 
directions  so that the net current is zero. For a sample with 
negative-charge carriers \textit{the (non-mesoscopic) current on the 
inner edge is counter-clockwise while that on the outer edge is clockwise}
as shown in Fig.~(\ref{fig:corbino}).
An intuitive way to determine these directions is provided by the 
\textit{skipping orbits} semi-classical picture: 
the electrons in the plane move along  \textit{clockwise} circles  
because the 
magnetic field points downwards. In the bulk the electron orbits are 
fixed complete circles so that there is no electron drift,  hence 
no net current. 
However on the inner edge the circular orbits are cut and elastically 
reflected from the edge becoming open curves resulting in a 
\textit{clockwise} drift of the inner-edge 
electrons producing a \textit{counter-clockwise} current $I_-$. 
On the outer edge the drift of the electrons in the cut orbits is 
\textit{counter-clockwise} so that the outer edge current $I_+$ is  
\textit{clockwise}. 
Quantum-mechanically the situation is quite 
similar only the effect of the edges is represented by a confining potential.
Note that the role of the confining potential is fairly important --
without it there would be no mesoscopic effects. 

Introducing adiabatically the AB  flux creates by Faraday induction  
an azimuthal electric field 
$\mathbf{E}=-\frac{1}{2\pi r} \frac{\d \phi}{\d t} \, \mathbf{e}_\varphi$.
which cannot create an azimuthal  bulk current since 
$\sigma_{xx}=0$ but induces a radial current pulse  due to the non-zero 
$\sigma_{xy}$
\[
{\bf i}=\nu_H \left[\begin{array}{rr} 0 & \ \ 1 \cr -1 & \ \ 0\end{array} 
\right] 
\mathbf{E}_{\phi}=
-\nu_H \frac{\d\phi}{\d t} \, \frac{1}{2\pi r} \, \mathbf{e}_r.
\]  
Basically the radial current pulse   creates a radial polarization 
which on its own induces an axial quasi-permanent electric current which 
is called persistent since it is observed for more than $10^3$ s in 
real samples. In more detail, 
the current pulse $\mathbf{i}$, corresponding to the Hall current 
which is purely bulk in this geometry, transfers electric charge
$\Delta Q= -e\nu_H\Delta\phi$ between the two edges  
which destroys the perfect balance between them. 
This pulse \textit{increases} the  carrier density
on the outer edge while \textit{decreasing} that on the inner one
and therefore \textit{increases} the outer-edge current 
while  \textit{decreasing} the inner-edge one by the same amount.
As a result  there appears a  net current which is  
diamagnetic, i.e., clockwise, 
for small positive $\Delta\phi$,   that is in the direction of the 
electric field
(under the reasonable assumption that the Fermi velocities do not 
change with the AB flux).
Note that the mesoscopic net current fluctuates with period $1$ and can 
therefore be also opposite to the electric field as the flux changes. 
The above argument also implies that 
\textit{the mesoscopic persistent currents 
(i.e., the changes in the edge currents due to the additional AB flux)
on both edges have the same direction (clockwise) and both 
produce negative magnetic moments}.

The typical form of the mesoscopic edge currents for $T=0$ 
can be obtained from the charge transferred between the edges:
\[
\Delta I=e v_F \Delta n = v_F \frac{\Delta Q}{L} = 
\pm\nu_H\frac{ev_F}{L}\, \phi,
\]
where we have divided $\Delta Q$ by the circumference $L$ in order to 
get the one-dimensional charge density on the edge 
($e\, L\, \Delta n=\Delta Q $). 
Thus we get
\[
I_{\out}=-\nu_H \frac{e v_F^{(\out)}}{2\pi R_{\out}} \,\phi, \quad 
I_{\inn}=- \nu_H \frac{e v_F^{(\inn)}}{2\pi R_{\inn}}\, \phi
\]
We note that the mesoscopic edge currents  are just small fluctuation 
to be added on top of the much bigger non-mesoscopic edge 
currents \cite{oscillate} produced by the huge background magnetic field. 
\subsection{The thermodynamic CFT derivation}
\label{sec:pers-CFT}
\begin{figure}[htb]
\centering
\caption{Persistent currents in the $k=2,3$ and $4$ parafermion FQH states,
as functions of the magnetic flux within one period,
computed numerically for $T/T_0=0.1$.
The flux is measured in units $\mathrm{h}/\e$ and the current's unit is
$\e v_F/L$.
The period is 1 flux quantum for all states and the amplitudes are
$I_2^{\max}=0.207$,  $I_3^{\max}=0.245$ and $I_4^{\max}=0.268$.
\label{fig:periods}}
\epsfig{file=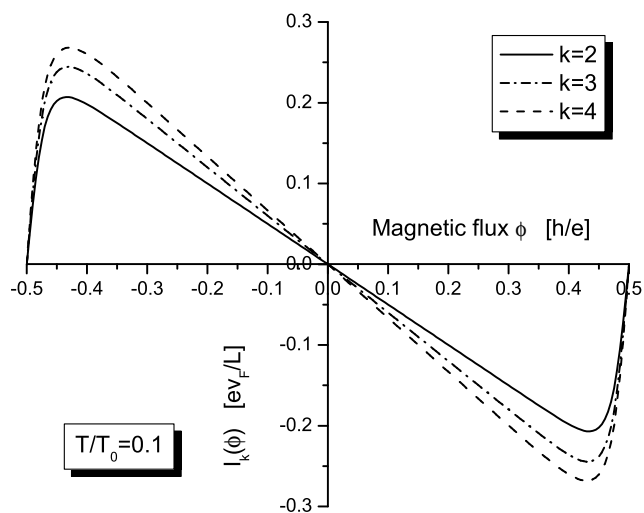,height=8cm}
\end{figure}

According to Eq.~(\ref{I}) and  the discussion in Sect.~\ref{sec:AB}, 
the equilibrium chiral persistent current in the FQH system
can be computed directly from the CFT partition function by
\beq\label{pers}
I=\left(\frac{e}{h}\right) k_B T \frac{\partial}{\partial \phi}
\ln \widetilde{Z}(\t,\phi\t) , \quad \mathrm{where}\quad
\t=i\pi \frac{T_0}{T}, \quad T_0=\frac{\hbar v_F}{\pi k_B L},
\eeq
$v_F$ is the Fermi velocity and $L=2\pi R$ the circumference of the
edge.
Because of Theorem 3 in \cite{byers-yang} the partition function
of the system should be an even periodic function of the flux $\phi$ with
period $1$. Therefore the persistent current (\ref{pers}) must be
an odd function with period $1$ so that it would be enough to
consider the half interval $-1/2 < \phi < 0$ and continue it by 
 $I(\phi)=- I(-\phi)$.
The $V$-invariance of the CFT partition function (\ref{Z_PF}) 
for the  $\PF_k$ states  
implies that the corresponding  chiral persistent current should be
periodic in $\phi$ with period at most  $1$. Any other value,
such as $1/2$, $1/3$ etc.,  of the persistent current's period
which is smaller than $1$ is a signal for some broken continuous
symmetry which is not allowed in unitary field theories in $1+1$
dimensions \cite{coleman,mermin-wagner}. Therefore we expect that
the period should be exactly one flux quantum.
The plots of the chiral persistent currents  in the first several
PF states computed numerically from Eq.~(\ref{pers})
for  $-1/2 \leq \phi \leq 1/2$ at temperature $T/T_0=0.1$, given on
Fig.~\ref{fig:periods},
\noindent
indicate that
\textit{all currents are periodic in $\phi$ with period exactly $1$}.
We  do not see any anomalous oscillations in the $k=2,3,4$ PF states, 
such as fractional flux periodicity of the persistent currents 
\cite{ino,ino2,kiryu} which are
characteristic for the BCS paired condensates, or more generally for some 
broken symmetries for all numerically accessible temperatures
$0.03\leq T/T_0 \leq 14$.
This is a confirmation  of the Bloch--Byers--Yang theorem \cite{byers-yang}
formulated  in the context of superconductors.
The persistent currents for the PF states   have been shown
to have oscillations  for $M=0$ \cite{kiryu}, however, this case is
physically irrelevant. The case $M=0$ has only been used in
\cite{rr} as a technical tool for investigating the wave functions
for the PF states. One of the objectives of this paper was to show
that the periods of the persistent currents for the PF states
are exactly $1$ for all temperature. Nevertheless, for $T=0$ it is
possible that the QH systems undergo phase transitions to some
BCS-like phases where the periods could be less then $1$ due to some
broken symmetries.

\begin{figure}[htb]
\centering
\caption{Temperature decay of the persistent current's amplitudes
in the $k=2,3$ and $4$ parafermion FQH states (without the contribution from
the two $\nu=1$ Landau levels with opposite spins)  computed numerically
in units $\e v_F/L$, for temperatures measured in units of $T_0$.
The zero temperature amplitude in these units are $\nu_k/2$, i.e.,
$1/4,3/10$ and $1/3$ respectively.
\label{fig:decay}}
\epsfig{file=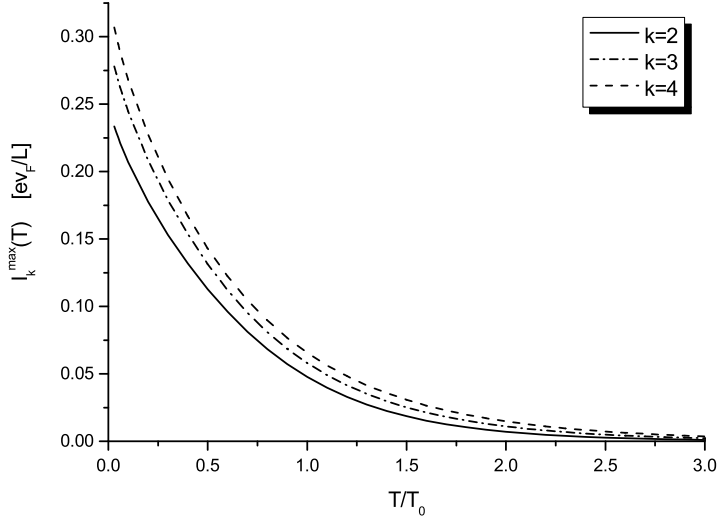,height=8cm}
\end{figure}

Furthermore, the amplitudes of the persistent currents in the 
parafermion FQH states decay exponentially with temperature, as shown
in Fig.~\ref{fig:decay}.

\subsection{Low-temperature asymptotics of the persistent currents
in the $\PF_k$ states}
\label{sec:low-T}
Because the low-temperature asymptotics of the chiral partition function
(\ref{Z_0}) includes only universal parameters such as 
$\nu_H$, $Q_{\mathrm{qh}}$ 
and the proper quasihole energy $\widetilde{\eps}_{\qh}$ 
the low-temperature behavior  of the persistent current 
which we analyze in this section would have a universal form and would 
be valid for all FQH states.

Taking the logarithm of Eq.~(\ref{Z_0}) and  differentiating with respect
to $\phi$ like in Eq.~(\ref{pers}) we get the following expression for
the  persistent current for $|\phi| < 1/2$
\beq\label{I-low}
I(T,\phi)  \  \mathop{=}_{T\ll T_0} \
\frac{\e v_F}{L}
\left\{ - \nu_H \phi +2 Q_\qh \e^{-\frac{\widetilde{\eps}_\qh}{k_B T}}
\sinh\left( 2 \pi^2 \frac{T_0}{T}Q_\qh\phi\right)    \right\}.
\eeq
\subsubsection{The persistent current's amplitude for $T=0$}
\label{sec:pers-T0}
For $T=0$ the second term in the curly brackets in Eq.~(\ref{I-low})
vanishes which expresses the fact that for zero temperature the free energy is
determined by the ground state energy. Then the persistent current is simply
proportional to the flux 
\[
I(T=0,\phi)=- \nu_H \,\frac{ev_F}{L}\, \phi,  \quad \mathrm{for}\quad 
-\frac{1}{2}\leq \phi\leq \frac{1}{2}, 
\]
which corresponds to the so called saw-tooth curve.  
It riches its maximum at $\phi=-1/2$ which is
\[
  I_{\max}=\frac{1}{2}\, \nu_H\, \frac{ev_F}{L}  \quad \mathrm{for} \ T=0 .
\]
We stress that the zero temperature behavior of the persistent currents 
is completely determined  by the CZ factors (\ref{CZ}).
\subsubsection{The persistent current's amplitude for $0<T\ll T_0$}
Because the persistent current is an odd periodic function
of the flux  we are going to consider the half-interval only
$-1/2 < \phi \leq 0$. Writing the $\sinh$ in Eq.~(\ref{I-low}) into its
 exponential form and ignoring $\exp(2\pi^2 T_0 Q_\qh \phi/T)$ which vanishes
 in the limit $T\to 0$ we get
 \[
 I(T,\phi) \  \mathop{=}_{T\ll T_0} \
\frac{\e v_F}{L}
\left\{ - \nu_H \phi - Q_\qh \e^{-\frac{\widetilde{\eps}_\qh}{k_B T}}
\e^{-2 \pi^2 \frac{T_0}{T}Q_\qh\phi}    \right\}.
 \]
The solution  of $\partial I(T,\phi)/\partial \phi =0$
for  $-1/2 < \phi \leq 0$ gives the position of the current's maximum
\beq\label{phi_max}
\phi_{\max}(T) \  \mathop{=}_{T\ll T_0} \
 - \frac{\D_\qh}{Q_\qh} +\frac{1}{2\pi^2 Q_\qh}\frac{T}{T_0}
\ln\left( \frac{2\pi^2Q_\qh^2}{\nu_H}\frac{T_0}{T}\right) 
\eeq
and substituting Eq.~(\ref{phi_max}) into Eq.~(\ref{I-low}) we get 
the following expression for the persistent current's amplitude
\beq\label{I_max}
I_{\max}(T)\  \mathop{=}_{T\ll T_0} \ \nu_H   \frac{\e v_F}{L}
\left\{  \frac{\D_\qh}{Q_\qh} -\frac{1}{2\pi^2 Q_\qh}\frac{T}{T_0}
	\left[1+\ln\left( \frac{2\pi^2 Q_\qh^2}{\nu_H}\frac{T_0}{T}\right) \right]
	\right\}.
\eeq
Using the charge--statistics relation Eq.~(\ref{c-s}) and the expression for 
the proper quasihole energy Eq.~(\ref{proper}) we can rewrite 
Eq.~(\ref{I_max}) in the form of Eq.~(\ref{I_max2}) 
according to which the persistent current's amplitude decays
logarithmically with increasing the temperature due to the thermal activation
of quasiparticle--quasihole pairs.
\subsection{High-temperature asymptotics of the persistent currents in 
the $\Z_k$ parafermion FQH states: extracting the universal decay 
exponents}
\label{sec:high-T}
\begin{figure}[htb]
\centering
\epsfig{file=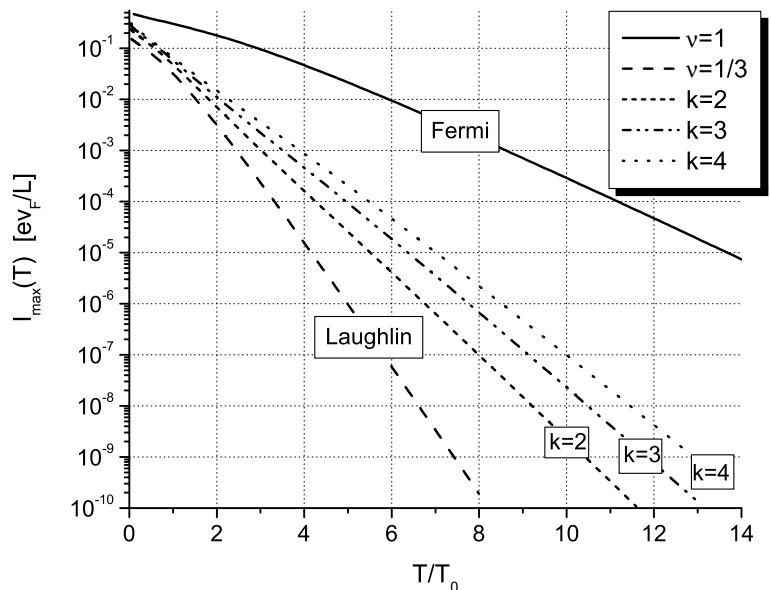,height=9cm}
\vspace{-1cm}
\caption{Logarithmic plot of the temperature decay of the persistent 
current's amplitudes for the Fermi liquid ($\nu_H=1$), the 
Laughlin state ($\nu_H=1/3$) and the $k=2,3$ and $4$ parafermion FQH 
states \label{fig:log-decay}}
\end{figure}

Unlike the analysis of the low-temperature asymptotics of the persistent 
currents in Sect.~\ref{sec:low-T} which has a universal form
depending only on the quasiparticle charges and proper quasihole 
energies the high-temperature asymptotics, which is also universal,
  depends on the explicit form of 
the modular $S$ matrix so in the next subsections we shall illustrate 
it on the examples of 
the $\Z_k$ parafermion FQH states for $k=2,3$ and $4$.
We show on Fig.~\ref{fig:log-decay} the logarithmic plot of the
persistent currents amplitudes for the Fermi liquid with $\nu=1$, 
the Laughlin state with $\nu=1/3$ and the parafermion FQH states 
with $k=2,3$, and $4$ computed numerically from the corresponding 
CFT partition functions for temperature $0.03\leq T/T_0\leq 14$
which includes also the high-temperature regime $T\gg T_0$.
\subsection{k=2}
According to Eq.~(\ref{l-rho})  the irreducible representations of the
 $\PF_2$ model are labelled by the pairs
$(l,\rho)$ where $-1\leq l \leq 2$ and $\rho=0,1$ with the restriction
$\rho \geq l-\rho \mod 2$. Using the explicit $S$ matrix for the Pfaffian 
FQH state from Ref.~\cite{CMP99} (see Eq.~(5.8) there) we find the
universal coefficients necessary for the computation of the high-temperature 
limit of the partition function~(\ref{Z-high})
\[
F^{l,\rho}=
\sum_{l'=-1}^2 \ \sum_{\rho'\geq l'-\rho'} S^{(l,\rho)}_{(l',\rho')}=
\left\{ \begin{array}{lcl}
	\sqrt{2}+1 & \mathrm{for} & l=0,\ \rho=0 \\
	\sqrt{2}-1 & \mathrm{for} & l=0,\ \rho=1 \\
	0 & & \mathrm{otherwise}
	\end{array}\right. .
\]
For convenience the characters of the $k=2$ parafermion FQH state are 
written explicitly 
in Eq.~(\ref{chi_PF_2}) and  the chiral partition function (\ref{Z_PF}) 
after $S$ transformation takes the form
\beqa
Z_2(T,\phi) &=& (\sqrt{2}+1) \chi_{0,0}(\t',\z') +
(\sqrt{2}-1) \chi_{0,1}(\t',\z') = \nn
&\simeq& \left(\sqrt{2}+1 \right)K_0(\t',2\z';8) +
 \left(\sqrt{2}-1 \right) K_4 (\t',2\z';8) ,
\eeqa
where we have dropped out a $\z'$-independent term proportional to the central
charge, substituted $\ch(\L_0+\L_0)(\t')\simeq {q'}^0$,
$\ch(\L_1+\L_1)(\t')\simeq {q'}^{1/2}$ and  ignored  ${q'}^{1/2}$ as compared
to   ${q'}^0$ since $q'\to 0$ when $T\to \infty$. Next, again dropping the
$\eta$ function and taking only the leading terms in the $K$ functions, i.e.,
$n=0$ for $K_0(8)$ and $n=0,-1$ for  $K_4(8)$  we get
\[
Z_2(T,\phi)  \ \mathop{\simeq}_{T\gg T_0}  \
\left(\sqrt{2}+1 \right)
\left(1+ 2\left( \frac{\sqrt{2}-1}{\sqrt{2}+1}\right) q' \cos(2\pi\z') \right).
\]
 Substituting the $T$ and $\phi$ according to Eq.~(\ref{new-param}) and
 Eq.~(\ref{q'}), using the approximation $\ln(1+X)\simeq X$ for $X\ll 1$
and taking the derivative with respect to $\phi$
 according to Eq.~(\ref{pers}) we get
 the following high-temperature asymptotic expression for the persistent
 current (note that $2\pi^2 k_B T_0/\phi_0= ev_F/L$)
 \beq\label{I_2}
	 I_2(T,\phi)   \mathop{\simeq}_{T\gg T_0}
	 - I_2^\max (T) \sin(2\pi \phi),   \quad
	  I_2^\max(T) = \frac{ev_F}{L} \frac{2}{\pi}
		\left(\frac{\sqrt{2}-1}{\sqrt{2}+1}\right)
   \frac{T}{T_0} \exp\left(-2 \frac{T}{T_0}\right)  \quad
 \eeq
\subsection{$k=3$}
For this case Eq.~(\ref{l-rho}) tells us that the irreducible 
representations are labelled by
$(l,\rho)$ where $-2\leq l \leq 2$ and $0\leq \rho\leq 2$ with the 
restriction $\rho \geq l-\rho \mod 3$. The $S$ matrix has been 
 written explicitly in Eq.~(5.7) in Ref.~\cite{NPB2001} and the 
sums of the $S$-matrix elements are
\[
F^{l,\rho}=\sum_{l'=-2}^2 \ \sum_{\rho'\geq l'-\rho'} 
S^{(l,\rho)}_{(l',\rho')}=
\left\{ \begin{array}{rcl}
  \frac{2}{\sqrt{5}} \left(\sin\left(\frac{\pi}{5}\right) +
3 \sin\left(\frac{2\pi}{5}\right) \right)  & \mathrm{for} & l=0,\ \rho=0 \\
  \frac{2}{\sqrt{5}} \left(3\sin\left(\frac{\pi}{5}\right) -
 \sin\left(\frac{2\pi}{5}\right) \right)  & \mathrm{for} & l=0,\ \rho=2 \\
	0 & & \mathrm{otherwise}
	\end{array}\right. .
\]
The 10 independent characters in this case are shown in
 Eq.~(\ref{chi_PF_3}) and the partition function (\ref{Z_PF}) 
for $k=3$ after $S$ transformation becomes
\beqa
 Z_3(T,\phi) &=& F^{0,0} \chi_{0,0}(\t',\z') +
F^{0,2} \chi_{0,2}(\t',\z') \simeq \nn
&\simeq& K_0(\t',3\z';15) \left( F^{0,0}{q'}^0+F^{0,2}{q'}^{2/5} \right) + \nn
 &+&  K_5(\t',3\z';15) \left( F^{0,0}{q'}^{2/3}+F^{0,2}{q'}^{1/15} \right) +  \nn
 &+&  K_{-5}(\t',3\z';15) \left( F^{0,0}{q'}^{2/3}+F^{0,2}{q'}^{1/15} \right),
\nonumber
\eeqa
where we have used $\ch(\L_0+\L_0)(\t')\simeq {q'}^0$,
$\ch(\L_0+\L_1)(\t')=\ch(\L_0+\L_2)(\t')\simeq {q'}^{1/15}$,
$\ch(\L_1+\L_1)(\t')=\ch(\L_2+\L_2)(\t')\simeq {q'}^{2/3}$  and
$\ch(\L_1+\L_2)(\t')\simeq {q'}^{2/5}$ (using Eq.~(\ref{D_PF}) for 
the CFT dimensions), again skipping the central charge
term in the characters (\ref{ch}). Keeping only the lowest powers in
$q'\to 0$, ignoring the Dedekind function and taking only $n=0$ in the
$K$ functions  we get
\[
Z_3(T,\phi) \ \mathop{\simeq}_{T\gg T_0}  \
F^{0,0}\left(1  + 2\frac{F^{0,2}}{F^{0,0}} \,
{q'}^{\frac{1}{2} \left[5/3 +2/15 \right]} \cos(2\pi\z')\right)
\]
To get the persistent current in the high-temperature limit we apply the
same scheme as for $k=2$ which yields
\beqa\label{I_3}
   I_3(T,\phi)   & \ \mathop{\simeq}_{T\gg T_0} \ &
   - I_3^\max(T) \sin(2\pi \phi),   \nn
    I_3^\max(T) & = & \frac{ev_F}{L} \frac{2}{\pi}
    \left(\frac{D-1}{D+1}\right)
   \frac{T}{T_0} \exp\left(- \frac{T}{T_0}\left[ \frac{5}{3} +
\frac{2}{15}\right]\right),  \quad
 \eeqa
where we have used that
\[
\frac{F^{0,2}}{F^{0,0}}= \frac{D-1}{D+1}, \quad \mathrm{where}
\quad D=2\cos\left(\frac{\pi}{5}\right)
\]
is the quantum dimension of the elementary quasihole in the $\PF_3$
FQH state \cite{NPB2001}.
\subsection{$k=4$}
For $k=4$ the representations are labelled by (see Eq.~(\ref{l-rho}))
$(l,\rho)$ where $-2\leq l \leq 3$ and $0\leq \rho\leq 3$ with the 
restriction $\rho \geq l-\rho \mod 4$.
The 10 dimensional $S$ matrix $\Sc$ for the (neutral) 
affine coset is shown in Eq.~(\ref{S_4}) and the complete 15 dimensional 
$S$ matrix is computed by Eq.~(\ref{S_PF}). Then the 
sums of the $S$-matrix elements are
\[
F^{l,\rho}=\sum_{l'=-2}^3 \ \sum_{\rho'\geq l'-\rho'} S^{(l,\rho)}_{(l',\rho')}=
\left\{ \begin{array}{rcl}
  2+\sqrt{3}  & \ \mathrm{for} \ & l=0,\ \rho=0 \\
  2-\sqrt{3} & \ \mathrm{for} \ & l=0,\ \rho=2 \\
   1 & \ \mathrm{for} \ & l=0,\ \rho=3 \\
  0 & & \ \mathrm{otherwise}
	\end{array}\right.
\]
so that the partition function (\ref{Z_PF}) for $k=3$ becomes
\beqa
 Z_4(T,\phi) &=& F^{0,0} \chi_{0,0}(\t',\z') +
F^{0,2} \chi_{0,2}(\t',\z') + F^{0,3} \chi_{0,3}(\t',\z')\simeq \nn
&\simeq& K_0(\t',4\z';24) \left(F^{0,0}{q'}^{\D^{\PF}(0,0)}+
F^{0,2}{q'}^{\D^{\PF}(2,2)} + F^{0,3}{q'}^{\D^{\PF}(1,3)}\right) + \nn
 &+&  K_6(\t',4\z';24) \left(F^{0,0}{q'}^{\D^{\PF}(1,1)}+
 F^{0,2}{q'}^{\D^{\PF}(3,3)} + F^{0,3}{q'}^{\D^{\PF}(0,2)}\right) + \nn
  &+&  K_{-6}(\t',4\z';24) \left(F^{0,0}{q'}^{\D^{\PF}(3,3)}+
 F^{0,2}{q'}^{\D^{\PF}(1,1)} + F^{0,3}{q'}^{\D^{\PF}(0,2)}\right),
\nonumber
\eeqa
where the complete characters are shown in Eq.~(\ref{chi_PF_4}) 
and  we have used
$\ch(\L_\mu+\L_\rho)(\t')\simeq {q'}^{\D^{\PF}(\mu,\rho)}$,
with
$\D^{\PF}(0,0)= 0 $, $\D^{\PF}(0,2)= 1/12 $,  
$\D^{\PF}(1,1)=\D^{\PF}(3,3)=3/4$,
$\D^{\PF}(1,3)=1/3$,  $\D^{\PF}(2,2)=1$ (see Eq.~(\ref{D_PF}) for the CFT 
dimensions $\D^{\PF}$),
again skipping the central charge
term in the characters (\ref{ch}). Keeping only the lowest powers in
$q'\to 0$, ignoring the Dedekind function and taking only $n=0$ in the
$K$ functions  we get
\[
Z_4(T,\phi) \ \mathop{\simeq}_{T\gg T_0}  \
F^{0,0}   + 2 F^{0,3}{q'}^{\frac{3}{4}+\D^{\PF}(0,2)}  \cos(2\pi\z')
\]
To get the persistent current in the high-temperature limit we apply the
same scheme as before
\beqa\label{I_4}
   I_4(T,\phi)   & \ \mathop{\simeq}_{T\gg T_0} \ &
   - I_4^\max (T) \sin(2\pi \phi),   \nn
    I_4^\max (T) & = & \frac{ev_F}{L} \frac{2}{\pi}
    \left(\frac{1}{2+\sqrt{3}}\right)
   \frac{T}{T_0} \exp\left(- \frac{T}{T_0}\left[ \frac{3}{2} +
\frac{2}{12}\right]\right).
\eeqa
 Finally, we can summarize the results for the high-temperature asymptotics
of the persistent current's  amplitude for the $\PF_k$ FQH states
 \beq\label{high-T}
I^{\max}_k(T) \ \mathop{\simeq}_{T\gg T_0} \ I_0^k \, 
\left(\frac{T}{T_0}\right) \, \exp\left(-\alpha_k\frac{T}{T_0}\right)
 \eeq
 where the exponent $\alpha_k$ for the $\Z_k$ parafermion FQH 
states is universal and can be written in general as
 \beq\label{alpha_k}
 \alpha_k = \frac{1}{\nu_H}+2\D^{\PF}\left(\L_0+\L_2\right),
\quad 2\D^{\PF}\left(\L_0+\L_2\right)=\frac{2(k-2)}{k(k+2)}.
 \eeq
 \begin{rem}\label{rem:alpha}
 While  $\alpha=1/\nu_H$ for the Fermi and Laughlin states 
\cite{geller-loss-kircz,geller-loss-kircz-1},  this  exponent is  not 
determined by the  filling factor alone in the $\Z_k$ parafermion FQH 
states.  
 Instead, there is a crucial  contribution from the neutral sector 
which is  described here by the  $\Z_k$ parafermion coset.
 \end{rem}
The analysis of the high-temperature asymptotics allows us to 
predict the form of the universal exponents $\alpha$
for  arbitrary FQH states. Because $F_\l\neq 0$  only if the 
corresponding electric charge $Q(\l)$ vanishes the neutral 
contribution to $\alpha$ would be
determined from the neutral weight $\L_{*}$ 
with minimal neutral CFT dimension which 
satisfies the pairing rule\footnote{Note that the characters
 $K_{\pm d_H}$ always appear in the vacuum character $\chi_{0,0}$  
together with $K_0$ and have the minimal nonzero CFT dimension} 
with $l=d_H$, i.e.,  
\beq\label{alpha}
\alpha=\frac{1}{\nu_H} + 2\D^{(0)}(\L_{*}), 
\quad   \mathrm{where} \quad
P[\L_{*}]\equiv d_H \mod k \quad   \mathrm{with} \quad
\D^{(0)}(\L_{*})=\min,
\eeq
i.e., $\L_*$ is the lowest  weight of the neutral character 
having the minimal CFT dimension among all neutral characters 
 which couple to the $\uu$ character 
$K_{\pm d_H}(\t,n_H\z;n_Hd_H)$.
Thus Eq.~(\ref{alpha_k})  should be valid for the 
entire parafermion hierarchy since $\L_*=\L_0+\L_2$
where $\D^{\PF}(\L_0+\L_2)=\min$ is the minimal dimension of the weights 
satisfying the pairing rule $P[\L_0+\L_2]=2 \equiv k+2 \mod k$.
 Accidentally, the parafermion CFT dimension $\D^\PF(\L_0+\L_2)=0$
  for $k=2$ (see Eq.~(\ref{D_PF})) and  the  exponent in that case 
is given by the	inverse of the filling factor only.

 Using MAPLE we have computed the values of $\alpha$ and
 $I_0$
 for the $k=2,3,4$ PF states
from the numerical data shown on Fig.~\ref{fig:log-decay}
 by linear regression
 (after subtracting the $\ln(T/T_0)$ contribution)  over 80 points
 in  the interval $6\leq T/T_0 \leq 14$.
 The exact values
 of the  exponents $\alpha$ and the (interpolated)   asymptotic
 amplitudes  $I_0$ for the first several $\PF_k$ states are in perfect 
agreement with the numerical ones and  both  are summarized in  
Table~\ref{tab:fit}. The exponents $\alpha_k$ for the parafermion FQH states 
have to be compared with those corresponding to Fermi and Laughlin liquids 
which according to Remark~\ref{rem:alpha} are
$\alpha=1$ and $\alpha=3$ respectively.
\begin{table}[htb]
	\centering
\caption{The slopes $\alpha$ and the $y$-intercepts $I_0$ (in units
		$\e v_F/L$) of the logarithmic plots in the high-temperature region
    for the $k=2,3,4$ PF states.
		\label{tab:fit}}
\begin{tabular}{|c||c|c||c|c|} \hline
  $k$ & \multicolumn{2}{c||}{$\alpha$} &
	\multicolumn{2}{c|}{$I_0$ $[\e v_F/L]$} \\
	\cline{2-5}
	& numeric& exact  & numeric & exact \\
	\hline\hline
	$2$ & $2.000706$ & $2$ &  $0.110253$ &
	$(2/\pi)(\sqrt{2}-1)/(\sqrt{2}+1)$ \\
  $3$ & $1.800010$ & $1.8$ & $0.150309$ & $(2/\pi)(2\cos(\pi/5)-1)/(2\cos(\pi/5)+1)$\\
	$4$ & $1.666306$ & $5/3$ & $0.169827$ & $(2/\pi)/(2+\sqrt{3})$\\
  \hline
\end{tabular}
\end{table}
As mentioned before, Eq.~(\ref{high-T}) demonstrates that at 
high temperature there exists a different mechanism reducing the 
persistent currents -- thermal decoherence 
which leads to exponential decay of the persistent currents amplitudes. 
\section{Conclusion}
We have shown that the two-dimensional CFT for the 
edge states of a FQH disk sample 
can be effectively used to compute analytically the partition function 
in presence of Aharonov--Bohm flux. 
The effect of the AB flux is interpreted as twisting of the electron 
operator and of the CFT Hamiltonian leading to the very compact formula
(\ref{Z_phi}) for the partition function in presence of flux.
Using this partition function as a 
thermodynamical potential one can compute the oscillating 
persistent current, AB magnetization and magnetic susceptibility 
in arbitrary FQH states. 
Because of the absence of complete spin--charge separation the 
excitations of the FQH liquid satisfy a $\Z_{n_H}$ pairing rule 
(\ref{Z_n}) that we have formulated in a general form.
As a result we find that in general the CFT characters should have 
a specific $\Z_{n_H}$ invariant structure (\ref{chi}) of sum of products of 
charged and neutral characters.
The general approach has been illustrated on 
the example of the parafermion FQH states in the second Landau level
using the results of Ref.~\cite{NPB2001} for the corresponding CFT. 
Our analysis shows that the persistent currents in the parafermion FQH states 
are odd periodic functions 
of the AB flux with period exactly one flux quantum as it should be 
according to the Bloch--Byers--Yang theorem.
We find that these currents are thermally suppressed by two different 
mechanisms in the low- and high- temperature limits, respectively and 
have universal non-Fermi liquid behavior. 

For the high-temperature asymptotics of the parafermion CFT we have 
computed explicitly the modular $S$ matrices (\ref{S_PF}) using the 
convenient 
quasiparticle basis of Ref.~\cite{NPB2001}. As a byproduct we have 
derived the $S$ matrix (\ref{Sc}) for the diagonal affine  
coset (\ref{PF_k}) in a very 
compact form and have proven that it is equivalent to the $S$ matrix of the
$\widehat{su(k)}_2$ current algebra, i.e., not only the irreducible 
representations of the coset (\ref{PF_k}) and $\widehat{su(k)}_2$ are 
labelled by the 
same weights but the fusion rules of both CFTs are identical.

The universal exponents (\ref{alpha_k})  extracted from the 
high-temperature asymptotics of the persistent currents 
show that for the FQH states with numerator of the filling factor $n_H>1$ 
there is always  neutral contribution depending on the neutral sector 
and the $\Z_{n_H}$ pairing rule. The experimental measurement of these 
exponents may be useful for the identification of the correct CFT 
describing a given  FQH universality class.
\begin{ack}
I would like to thank Andrea Cappelli, Ivan Todorov, Christoph 
Schweigert and Michael Geller for useful discussions.
This work has been partially supported by the FP5-EUCLID Network Program
 of the European Commission under Contract No. HPRN-CT-2002-00325
and by the Bulgarian National
Council for Scientific Research under Contract No. F-1401.
\end{ack}
\begin{appendix}
\section{Chiral partition functions for the parafermion QH states}
For $k=2$ the $6$ independent characters entering Eq.~(\ref{Z_PF}) 
using the notation 
$K_{l} \left( 8 \right):=K_{l} \left(\t,2\z; 8 \right) $   and
$\ch(\mu,\rho):= \ch\left(\underline{\Lambda}_{\mu}+\underline{\Lambda}_{\rho} \right)(\t)$ with $0\leq\mu\leq\rho\leq 1$ can be written 
\beqa\label{chi_PF_2}
\chi_{-2,0}(\t,\z) &=& K_{{-2}} \left( 8 \right)
\ch \left( {0},{0} \right) +K_{{2}} \left( 8 \right) 
\ch \left({1},{1} \right)
\nn
\chi_{-2,1}(\t,\z) &=& K_{{-2}} \left( 8 \right)
\ch \left( {{1}},{{1}} \right) +
K_{{2}} \left( 8 \right) \ch \left( {0},{0} \right)
\nn
\chi_{-1,1}(\t,\z) &=& K_{{-1}} \left( 8 \right)
\ch \left({0},{1} \right) +
K_{{3}} \left( 8 \right) \ch \left( {0},{1} \right)
\nn
\chi_{0,0}(\t,\z) &=& K_{{0}} \left( 8 \right) \ch
 \left({0},{0} \right) +
K_{{4}} \left( 8 \right) \ch \left({1},{1} \right)
\nn
\chi_{0,1}(\t,\z) &=& K_{{0}} \left( 8 \right) \ch
 \left({1},{1} \right) +
K_{{4}} \left( 8 \right) \ch \left({0},{0} \right)
\nn
\chi_{1,1}(\t,\z) &=& K_{{1}} \left( 8 \right) \ch
 \left( {0},{1} \right) +
K_{{-3}} \left(8 \right) \ch \left({0},{1} \right)
\eeqa
For the $k=3$ parafermion state we use the following notation:
$K_{l} \left( 15 \right):=K_{l} \left(\t,3\z; 15 \right) $   and
$\ch(\mu,\rho):=
\ch\left(\underline{\Lambda}_{\mu}+\underline{\Lambda}_{\rho} \right)(\t)$
with $0\leq\mu\leq\rho\leq 2$ the $10$ inequivalent characters are
\begin{eqnarray}\label{chi_PF_3}
\chi_{-2,1} &=& K_{{-2}} \left( 15 \right) {\mathrm{ch}} \left( 0,1 \right) +K_{{3}}
 \left( 15 \right) {\mathrm{ch}} \left( 1,2 \right) +K_{{-7}} \left( 15
 \right) {\mathrm{ch}} \left( 0,2 \right)\nn
\chi_{-2,2} &=& K_{{-2}} \left( 15 \right) {\mathrm{ch}} \left( 2,2 \right) +K_{{3}}
 \left( 15 \right) {\mathrm{ch}} \left( 0,0 \right) +K_{{-7}} \left( 15
 \right) {\mathrm{ch}} \left( 1,1 \right) \nn
\chi_{-1,1} &=& K_{{-1}} \left( 15 \right) {\mathrm{ch}} \left( 0,2 \right) +K_{{4}}
 \left( 15 \right) {\mathrm{ch}} \left( 0,1 \right) +K_{{-6}} \left( 15
 \right) {\mathrm{ch}} \left( 1,2 \right) \nn
\chi_{-1,2} &=& K_{{-1}} \left( 15 \right) {\mathrm{ch}} \left( 1,1 \right) +K_{{4}}
 \left( 15 \right) {\mathrm{ch}} \left( 2,2 \right) +K_{{-6}} \left( 15
 \right) {\mathrm{ch}} \left( 0,0 \right) \nn
\chi_{0,0} &=& K_{{0}} \left( 15 \right) {\mathrm{ch}} \left( 0,0 \right) +K_{{5}} \left(
15 \right) {\mathrm{ch}} \left( 1,1 \right) +K_{{-5}} \left( 15 \right) {
\mathrm{ch}} \left( 2,2 \right)\nn
\chi_{0,2} &=& K_{{0}} \left( 15 \right) {\mathrm{ch}} \left( 1,2 \right) +K_{{5}} \left(
15 \right) {\mathrm{ch}} \left( 0,2 \right) +K_{{-5}} \left( 15 \right) {
\mathrm{ch}} \left( 0,1 \right) \nn
\chi_{1,1} &=& K_{{1}} \left( 15 \right) {\mathrm{ch}} \left( 0,1 \right) +K_{{6}} \left(
15 \right) {\mathrm{ch}} \left( 1,2 \right) +K_{{-4}} \left( 15 \right) {
\mathrm{ch}} \left( 0,2 \right) \nn
\chi_{1,2} &=& K_{{1}} \left( 15 \right) {\mathrm{ch}} \left( 2,2 \right) +K_{{6}} \left(
15 \right) {\mathrm{ch}} \left( 0,0 \right) +K_{{-4}} \left( 15 \right) {
\mathrm{ch}} \left( 1,1 \right) \nn
\chi_{2,1} &=& K_{{2}} \left( 15 \right) {\mathrm{ch}} \left( 0,2 \right) +K_{{7}} \left(
15 \right) {\mathrm{ch}} \left( 0,1 \right) +K_{{-3}} \left( 15 \right) {
\mathrm{ch}} \left( 1,2 \right) \nn
\chi_{2,2} &=& K_{{2}} \left( 15 \right) {\mathrm{ch}} \left( 1,1 \right) +K_{{7}} \left(
15 \right) {\mathrm{ch}} \left( 2,2 \right) +K_{{-3}} \left( 15 \right) {
\mathrm{ch}} \left( 0,0 \right)
\end{eqnarray}

For $k=4$ parafermion state  where 
$K_{l} \left( 24 \right):=K_{l} \left(\t,4\z; 24 \right) $   and
$\ch(\mu,\rho):=
\ch\left(\underline{\Lambda}_{\mu}+\underline{\Lambda}_{\rho} \right)(\t)$
with $0\leq\mu\leq\rho\leq 3$ we have $15$ independent characters
\begin{eqnarray}\label{chi_PF_4}
\chi_{-3,1} &=& K_{{-3}} \left( 24 \right) \mathrm{ ch} \left( 0,1 \right) +K_{{3}}
 \left( 24 \right) \mathrm{ ch} \left( 1,2 \right) +K_{{9}} \left( 24
 \right) \mathrm{ ch} \left( 2,3 \right) +K_{{-9}} \left( 24 \right) \mathrm{
ch} \left( 0,3 \right) \nn
\chi_{-3,3} &=& K_{{-3}} \left( 24 \right) \mathrm{ ch} \left( 2,3 \right) +K_{{3}}
 \left( 24 \right) \mathrm{ ch} \left( 0,3 \right) +K_{{9}} \left( 24
 \right) \mathrm{ ch} \left( 0,1 \right) +K_{{-9}} \left( 24 \right) \mathrm{
ch} \left( 1,2 \right) \nn
\chi_{-2,1} &=& K_{{-2}} \left( 24 \right) \mathrm{ ch} \left( 0,2 \right) +K_{{4}}
 \left( 24 \right) \mathrm{ ch} \left( 1,3 \right) +K_{{10}} \left( 24
 \right) \mathrm{ ch} \left( 0,2 \right) +K_{{-8}} \left( 24 \right) \mathrm{
ch} \left( 1,3 \right) \nn
\chi_{-2,2} &=& K_{{-2}} \left( 24 \right) \mathrm{ ch} \left( 1,1 \right) +K_{{4}}
 \left( 24 \right) \mathrm{ ch} \left( 2,2 \right) +K_{{10}} \left( 24
 \right) \mathrm{ ch} \left( 3,3 \right) +K_{{-8}} \left( 24 \right) \mathrm{
ch} \left( 0,0 \right) \nn
\chi_{-2,3} &=& K_{{-2}} \left( 24 \right) \mathrm{ ch} \left( 3,3 \right) +K_{{4}}
 \left( 24 \right) \mathrm{ ch} \left( 0,0 \right) +K_{{10}} \left( 24
 \right) \mathrm{ ch} \left( 1,1 \right) +K_{{-8}} \left( 24 \right) \mathrm{
ch} \left( 2,2 \right)    \nn
\chi_{-1,2} &=& K_{{-1}} \left( 24 \right) \mathrm{ ch} \left( 0,3 \right) +K_{{5}}
 \left( 24 \right) \mathrm{ ch} \left( 0,1 \right) +K_{{11}} \left( 24
 \right) \mathrm{ ch} \left( 1,2 \right) +K_{{-7}} \left( 24 \right) \mathrm{
ch} \left( 2,3 \right) \nn
\chi_{-1,3} &=& K_{{-1}} \left( 24 \right) \mathrm{ ch} \left( 1,2 \right) +K_{{5}}
 \left( 24 \right) \mathrm{ ch} \left( 2,3 \right) +K_{{11}} \left( 24
 \right) \mathrm{ ch} \left( 0,3 \right) +K_{{-7}} \left( 24 \right) \mathrm{
ch} \left( 0,1 \right)    \nn
\chi_{0,0} &=& K_{{0}} \left( 24 \right) \mathrm{ ch} \left( 0,0 \right) +K_{{6}} \left(
24 \right) \mathrm{ ch} \left( 1,1 \right) +K_{{12}} \left( 24 \right) {
\mathrm{ch}} \left( 2,2 \right) +K_{{-6}} \left( 24 \right) \mathrm{ ch}
 \left( 3,3 \right) \nn
\chi_{0,2} &=& K_{{0}} \left( 24 \right) \mathrm{ ch} \left( 1,3 \right) +K_{{6}} \left(
24 \right) \mathrm{ ch} \left( 0,2 \right) +K_{{12}} \left( 24 \right) {
\mathrm{ch}} \left( 1,3 \right) +K_{{-6}} \left( 24 \right) \mathrm{ ch}
 \left( 0,2 \right) \nn
\chi_{0,3} &=& K_{{0}} \left( 24 \right) \mathrm{ ch} \left( 2,2 \right) +K_{{6}} \left(
24 \right) \mathrm{ ch} \left( 3,3 \right) +K_{{12}} \left( 24 \right) {
\mathrm{ch}} \left( 0,0 \right) +K_{{-6}} \left( 24 \right) \mathrm{ ch}
 \left( 1,1 \right) \nn
\chi_{1,1} &=& K_{{1}} \left( 24 \right) \mathrm{ ch} \left( 0,1 \right) +K_{{7}} \left(
24 \right) \mathrm{ ch} \left( 1,2 \right) +K_{{-11}} \left( 24 \right) {
\mathrm{ch}} \left( 2,3 \right) +K_{{-5}} \left( 24 \right) \mathrm{ ch}
 \left( 0,3 \right) \nn
\chi_{1,3} &=& K_{{1}} \left( 24 \right) \mathrm{ ch} \left( 2,3 \right) +K_{{7}} \left(
24 \right) \mathrm{ ch} \left( 0,3 \right) +K_{{-11}} \left( 24 \right) {
\mathrm{ch}} \left( 0,1 \right) +K_{{-5}} \left( 24 \right) \mathrm{ ch}
 \left( 1,2 \right) \nn
\chi_{2,1} &=& K_{{2}} \left( 24 \right) \mathrm{ ch} \left( 0,2 \right) +K_{{8}} \left(
24 \right) \mathrm{ ch} \left( 1,3 \right) +K_{{-10}} \left( 24 \right) {
\mathrm{ch}} \left( 0,2 \right) +K_{{-4}} \left( 24 \right) \mathrm{ ch}
 \left( 1,3 \right) \nn
\chi_{2,2} &=& K_{{2}} \left( 24 \right) \mathrm{ ch} \left( 1,1 \right) +K_{{8}} \left(
24 \right) \mathrm{ ch} \left( 2,2 \right) +K_{{-10}} \left( 24 \right) {
\mathrm{ch}} \left( 3,3 \right) +K_{{-4}} \left( 24 \right) \mathrm{ ch}
 \left( 0,0 \right) \nn
\chi_{2,3} &=& K_{{2}} \left( 24 \right) \mathrm{ ch} \left( 3,3 \right) +K_{{8}} \left(
24 \right) \mathrm{ ch} \left( 0,0 \right) +K_{{-10}} \left( 24 \right) {
\mathrm{ch}} \left( 1,1 \right) +K_{{-4}} \left( 24 \right) \mathrm{ ch}
 \left( 2,2 \right)\nn
\end{eqnarray}
\section{The diagonal coset's  $S$ matrix}
\label{app:coset_S}
In this appendix we shall sketch the  derivation of
the modular $S$ matrix for the  coset  (\ref{PF_k}) following 
the general approach of Goddard--Kent--Olive \cite{gko}. 
The characters of the diagonal coset appear by definition 
as string functions in the decomposition of the coset numerator
$\widehat{su(k)}_1\oplus \widehat{su(k)}_1$ in terms of the 
$\widehat{su(k)}_2$ characters
\beq\label{decomp}
\chi_{\Lambda}^{(1)}(\t) \ \chi_{\Lambda'}^{(1)}(\t)=
\sum_{\Lambda''}\ P(\Lambda,\Lambda'; \Lambda'') \
\ch{\scriptsize\left(\begin{array}{c} \Lambda \ , \ \Lambda' \\ 
\Lambda'' \end{array} \right)}(\t) 
\ \chi_{\Lambda''}^{(2)}(\t),
\eeq
where $\chi_{\Lambda}^{(1)}$ and $\chi_{\Lambda}^{(2)}$ are the
$\widehat{su(k)}_l$ characters at  level $l=1$ and $2$ respectively
and the function
\[
P(\Lambda,\Lambda'; \Lambda'')= \left\{
\begin{array}{cl} 
1 &  \quad \mathrm{if}\quad [\Lambda'']=[\Lambda]+[\Lambda'] \\
0 &  \quad \mathrm{otherwise} 
\end{array}\right.
\]
expresses the $\Z_k$ charge conservation rule \cite{schw}.
Let us now apply the $S$ transformation $\t\to -1/\t$ to both sides, i.e., 
$S\chi_{\Lambda}^{(1)}=\sum_{\mu}S^{(1)}_{\Lambda,\mu}\chi_{\mu}^{(1)}$,
$S\chi_{\Lambda'}^{(1)}=\sum_{\mu'}S^{(1)}_{\Lambda',\mu'}\chi_{\mu'}^{(1)}$
$S\chi_{\Lambda''}^{(2)}=\sum_{\sigma}S^{(2)}_{\Lambda'',\sigma}
\chi_{\sigma}^{(2)}$ and 
$S\ch_{\rho''}=\sum_{\rho''}\Sc^{\Lambda''}_{\rho''}\ch_{\rho''}$
and then decompose again the product 
$\chi_{\mu}^{(1)}(\t)\chi_{\mu'}^{(1)}(\t)$ in the left hand side
in terms of the $\chi_{\mu''}^{(2)}$ and the coset characters
using Eq.~(\ref{decomp})
\beqa
&&\sum_{\mu,\mu' \mod k} \ \sum_{\mu''} P(\mu,\mu'; \mu'')  
S^{(1)}_{\Lambda,\mu} S^{(1)}_{\Lambda',\mu'} \  
\ch\left(\scriptsize\begin{array}{c} \mu \ , \ \mu' \\ 
\mu'' \end{array} \right)(\t) \ 
\chi_{\mu''}^{(2)}(\t) = \nn
&&=\sum_{\Lambda''} P(\Lambda,\Lambda'; \Lambda'') \sum_{\rho''} 
\sum_{\sigma} 
 \Sc^{\Lambda''}_{\rho''} S^{(2)}_{\Lambda'',\sigma} 
\ch\left(\scriptsize\begin{array}{c} [\rho''] \ , \ 0 \\ 
\rho'' \end{array} \right)(\t) 
\ \chi_{\sigma}^{(2)}(\t), \nonumber
\eeqa
where we have used the triple equivalence to label the coset
representations by the level-2 index only. Next we substitute 
$\sigma=\mu''$ in the RHS,  compare the coefficients of 
$\chi_{\mu''}^{(2)}$,  multiply both sides by $(S^{(2)}_{\mu'',\nu})^{-1}$
and sum over $\mu''$ to get
\beqa
&&\sum_{\mu,\mu' \mod k} \ \sum_{\mu''} P(\mu,\mu'; \mu'')  
S^{(1)}_{\Lambda,\mu} S^{(1)}_{\Lambda',\mu'} \   
(S^{(2)}_{\mu'',\nu})^{-1} 
\ch\left(\scriptsize\begin{array}{c} \mu \ , \ \mu' \\ 
\mu'' \end{array} \right)(\t)= \nn
&&= P(\Lambda,\Lambda'; \nu) \sum_{\rho''} 
 \Sc^{\nu}_{\rho''}  
\ch\left(\scriptsize\begin{array}{c} [\rho''] \ , \ 0 \\ 
\rho'' \end{array} \right)(\t). \nonumber
\eeqa
Because 
$(\mu,\mu';\mu'')\df (\L_{\mu},\L_{\mu'};\L_{\alpha}+\L_{\beta} )$ we can 
shift the summation variables in the LHS according to
\[
\mu=\widetilde{\mu}+\mu',\quad \alpha=\widetilde{\alpha}+\mu',
\quad \beta=\widetilde{\beta}+\mu'
\]
and use the triples equivalence 
$(\L_{\mu},\L_{\mu'};\L_{\alpha}+\L_{\beta} )\equiv 
(\L_{\widetilde{\mu}},0;\L_{\widetilde{\alpha}}+\L_{\widetilde{\beta}} )$
to get for the LHS
\[
\sum_{\mu' \mod k} \ \sum_{\widetilde{\mu} \mod k} 
\sum_{\widetilde{\alpha}\leq \widetilde{\beta}} 
P(\widetilde{\mu},0; \widetilde{\alpha}+\widetilde{\beta}) \, 
S^{(1)}_{\Lambda,\widetilde{\mu}+\mu'} S^{(1)}_{\Lambda',\mu'} \   
(S^{(2)}_{\widetilde{\alpha}+\mu'+\widetilde{\beta}+\mu',\nu})^{-1} \,
\ch\left(\scriptsize\begin{array}{c} \widetilde{\mu} \ , \ 0 \\ 
\widetilde{\alpha}+\widetilde{\beta} \end{array} \right)(\t),
\]
where we have used that 
$P(\widetilde{\mu}+\mu',\mu'; \widetilde{\alpha}+\mu'+\widetilde{\beta}+\mu')=
P(\widetilde{\mu},0; \widetilde{\alpha}+\widetilde{\beta})$.
Now we can use the properties of the $S$ matrices under the action of 
simple currents 
\[
S^{(1)}_{\Lambda,\widetilde{\mu}+\mu'}=\e^{2\pi i\frac{\mu'\Lambda}{k}}
S^{(1)}_{\Lambda,\widetilde{\mu}},\quad
S^{(1)}_{\Lambda',\mu'}=\e^{2\pi i\frac{\mu'\Lambda'}{k}}
S^{(1)}_{\Lambda',0},\quad
S^{(2)}_{\Lambda_{\widetilde{\alpha}+\mu'}+ \Lambda_{\widetilde{\beta}+\mu'}}
=\e^{-2\pi i Q^{(2)}_{\J^{\mu'}}(\nu)}
S^{(2)}_{\Lambda_{\widetilde{\alpha}}+ \Lambda_{\widetilde{\beta}},\nu},
\]
where $Q^{(2)}_{\J^{\mu'}}(\nu)$ is the monodromy charge (\ref{simp_prop}) 
in the $\widehat{su(k)}_2$ CFT. Writing the $\widehat{su(k)}_2$ weight
$\nu=\L_{\rho}+\L_{\sigma}$ with $0\leq\rho\leq \sigma\leq k-1$ 
and using the following formula for the 
CFT dimensions in $\widehat{su(k)}_2$
\[
\Delta^{(2)}(\L_{\rho}+\L_{\sigma})=
\frac{2\rho(k-\sigma)+(k+1)\left[ \rho(k-\rho)+\sigma(k-\sigma) \right]}{2k(k+2)}
\]
we get the following simple expression for the monodromy charge
\[
Q^{(2)}_{\J^{\mu'}}(\L_{\rho}+\L_{\sigma})=-\frac{\mu'(\rho+\sigma)}{k}.
\]
The final step is to take the summation over $\mu'$ using again the 
$\delta$-function \footnote{note that 
the $k$-ality of $\nu$ is $[\nu]=[\L_{\rho}+\L_{\sigma}]=\rho+\sigma$}
\[
\sum_{\mu' \mod k} 
\exp\left(2\pi i\, \frac{\mu'\left[ \Lambda+\Lambda'-(\rho+\sigma)\right]}{k}\right) 
= k \, P( \Lambda,\Lambda';\nu),
\]
set $\rho''=\Lambda_{\widetilde{\alpha}}+ \Lambda_{\widetilde{\beta}}$
(then $\widetilde{\mu}=\L_{[\rho'']}=:[\rho'']$ due to the conservation of 
the $\Z_k$ charge in the coset) and compare the 
coefficients of 
$\ch{\scriptsize\left(\begin{array}{c} [\rho''] \ , \ 0 \\ 
\rho'' \end{array} \right)}(\t)$ which gives the coset $S$ matrix, 
Eq.~(\ref{Sc}).

For the high-temperature asymptotics of the persistent current in the 
$k=4$ pa\-ra\-fer\-mi\-on FQH state we shall need the (neutral) $S$ matrix 
(\ref{Sc}). 
Using the following basis of $10$ coset weights
\beqa
&\L_0+\L_0,& \quad \L_0+\L_1, \quad \L_0+\L_2, \quad \L_0+\L_3, \quad 
\L_1+\L_1,   \nn
&\L_1+\L_2,&  \quad \L_1+\L_3, \quad \L_2+\L_2, \quad \L_2+\L_3, \quad 
\L_3+\L_3,  \nonumber
\eeqa 
the matrix $\Sc^{\mu,\nu}_{\rho,\sigma}$ has been computed with 
Maple by Eq.~(\ref{Sc}) where the $S$ matrix of the denominator 
 $\widehat{su(4)}_2$  of the coset has been computed by the Kac 
formula (\ref{Kac})  
\beq\label{S_4}
\left[\Sc^{\mu,\nu}_{\rho,\sigma}\right]=
\left[\begin{array}{cccccccccc} 
\frac{\sqrt{6}}{12} & \frac{\sqrt{2}}{4} & \frac{\sqrt{6}}{6} &
\frac{\sqrt{2}}{4} & \frac{\sqrt{6}}{12} & \frac{\sqrt{2}}{4} &
\frac{\sqrt{6}}{6} & \frac{\sqrt{6}}{12} & \frac{\sqrt{2}}{4} &
\frac{\sqrt{6}}{12} \\
\frac{\sqrt{2}}{4} & \frac{1+i}{4} & 0 & \frac{1-i}{4} &
\frac{i\sqrt{2}}{4}& \frac{-1+i}{4} & 0 & \frac{-\sqrt{2}}{4} & 
\frac{-1-i}{4} & \frac{-i\sqrt{2}}{4} \\
\frac{\sqrt{6}}{6} & 0 & \frac{\sqrt{6}}{6} & 0 &
\frac{-\sqrt{6}}{6} & 0 & \frac{-\sqrt{6}}{6} & \frac{\sqrt{6}}{6} &
0 & \frac{-\sqrt{6}}{6} \\
\frac{\sqrt{2}}{4} & \frac{1-i}{4} & 0 & \frac{1+i}{4} & 
\frac{-i\sqrt{2}}{4} & \frac{-1-i}{4} & 0 & \frac{-\sqrt{2}}{4} &
\frac{-1+i}{4} & \frac{i\sqrt{2}}{4} \\
\frac{\sqrt{6}}{12} & \frac{i\sqrt{2}}{4} & \frac{-\sqrt{6}}{6} &
\frac{-i\sqrt{2}}{4} & \frac{-\sqrt{6}}{12} & \frac{-i\sqrt{2}}{4} &
\frac{\sqrt{6}}{6} & \frac{\sqrt{6}}{12} & \frac{i\sqrt{2}}{4} &
\frac{-\sqrt{6}}{12}\\
\frac{\sqrt{2}}{4} & \frac{-1+i}{4} & 0 & \frac{-1-i}{4} & 
 \frac{-i\sqrt{2}}{4} & \frac{1+i}{4} & 0 & \frac{-\sqrt{2}}{4} &
\frac{1-i}{4} & \frac{i\sqrt{2}}{4} \\
\frac{\sqrt{6}}{6} & 0 & \frac{-\sqrt{6}}{6} & 0 & \frac{\sqrt{6}}{6} & 
0  & \frac{-\sqrt{6}}{6} & \frac{\sqrt{6}}{6} & 0 & 
\frac{\sqrt{6}}{6}\\
\frac{\sqrt{6}}{12} & \frac{-\sqrt{2}}{4} & \frac{\sqrt{6}}{6} & 
\frac{-\sqrt{2}}{4} & \frac{\sqrt{6}}{12} & \frac{-\sqrt{2}}{4} & 
\frac{\sqrt{6}}{6} & \frac{\sqrt{6}}{12} & \frac{-\sqrt{2}}{4} & 
\frac{\sqrt{6}}{12} \\
\frac{\sqrt{2}}{4} &  \frac{-1-i}{4} & 0 &  \frac{-1+i}{4} & 
\frac{i\sqrt{2}}{4} & \frac{1-i}{4} & 0 & \frac{-\sqrt{2}}{4} & 
\frac{1+i}{4} & \frac{-i\sqrt{2}}{4} \\
\frac{\sqrt{6}}{12} & \frac{-i\sqrt{2}}{4} & \frac{-\sqrt{6}}{6} & 
\frac{i\sqrt{2}}{4} & \frac{-\sqrt{6}}{12} & \frac{i\sqrt{2}}{4} & 
\frac{\sqrt{6}}{6} & \frac{\sqrt{6}}{12} & \frac{-i\sqrt{2}}{4} & 
\frac{-\sqrt{6}}{12} \\
\end{array} \right]
\eeq

\end{appendix}

\def\NP{Nucl. Phys. }
\def\PRL{Phys. Rev. Lett.}
\def\PL{Phys. Lett. }
\def\PR{Phys. Rev. }
\def\CMP{Commun. Math. Phys. }
\def\IJMP{Int. J. Mod. Phys. }
\def\JSP{J. Stat. Phys. }
\def\JP{J. Phys. }
\bibliography{Z_k,my}

\providecommand{\href}[2]{#2}\begingroup\raggedright\begin{thebibliography}{10}

\bibitem{pers-exp}
D.~Mailly, C.~Chapelier, and A.~Benoit, ``Experimental observation of
  persistent currents in a {G}a{A}s-{A}l{G}a{A}s single loop,'' {\em Phys. Rev.
  Lett.} {\bf 70} (1993) 2020.

\bibitem{fro-stu-thi}
J.~Fr\"{o}hlich, U.~M. Studer, and E.~Thiran, ``A classification of quantum
  {H}all fluids,'' {\em J. Stat. Phys.} {\bf 86} (1997) 821,
  \href{http://xxx.lanl.gov/abs/cond-mat/9503113}{{\tt cond-mat/9503113}}.

\bibitem{geller-loss-kircz}
M.~Geller, D.~Loss, and G.~Kirczenow, ``Mesoscopic effects in the fractional
  quantum {H}all regime: chiral {L}uttinger liquid versus {F}ermi liquid,''
  {\em Phys. Rev. Lett.} {\bf 77} (1996) 49.

\bibitem{geller-loss}
M.~Geller and D.~Loss, ``Aharonov--{B}ohm effect in the chiral {L}uttinger
  liquid,'' {\em Phys. Rev. B} {\bf 56} (1997) 9692.

\bibitem{grayson}
M.~Grayson, D.~Tsui, L.~Pfeiffer, K.~West, and A.~Chang, ``Continuum of chiral
  {L}uttinger liquids at the fractional quantum {H}all edge,'' {\em Phys. Rev.
  Lett.} {\bf 80} (1998) 1062.

\bibitem{5-2}
L.~Georgiev, ``The $\nu=5/2$ quantum {H}all state revisited: spontaneous
  breaking of the chiral fermion parity and phase transition between abelian
  and non-abelian statistics,'' {\em Nucl. Phys.} {\bf B 651} (2003) 331--360,
  \href{http://xxx.lanl.gov/abs/hep-th/0108173}{{\tt hep-th/0108173}}.

\bibitem{PRB-PF_k}
L.~Georgiev, ``Chiral persistent currents and magnetic susceptibilities in the
  parafermion quantum {H}all states in the second {L}andau level with
  {A}haronov--{B}ohm flux,'' {\em Phys. Rev.} {\bf B 69} (2004) 085305,
  \href{http://xxx.lanl.gov/abs/cond-mat/0311339}{{\tt cond-mat/0311339}}.

\bibitem{NPB2001}
A.~Cappelli, L.~Georgiev, and I.~Todorov, ``Parafermion {H}all states from
  coset projections of abelian conformal theories,'' {\em Nucl. Phys.} {\bf B
  599 [FS]} (2001) 499--530, \href{http://xxx.lanl.gov/abs/hep-th/0009229}{{\tt
  hep-th/0009229}}.

\bibitem{oscillate}
L.~Georgiev and M.~Geller, ``Magnetic moment oscillations in a quantum {H}all
  ring,'' (2004) \href{http://xxx.lanl.gov/abs/cond-mat/0404681}{{\tt
  cond-mat/0404681}}.

\bibitem{ino}
K.~Ino, ``Pairing effects at the edge of paired quantum {H}all states,'' {\em
  Phys. Rev. Lett.} {\bf 81} (1998) 1078,
  \href{http://xxx.lanl.gov/abs/cond-mat/9803337}{{\tt cond-mat/9803337}}.

\bibitem{fro2000}
J.~Fr\"ohlich, B.~Pedrini, C.~Schweigert, and J.~Walcher, ``Universality in
  quantum {H}all systems: Coset construction of incompressible states,'' {\em
  J. Stat. Phys.} {\bf 103} (2001) 527,
\href{http://xxx.lanl.gov/abs/cond-mat/0002330}{{\tt cond-mat/0002330}}.

\bibitem{wen-top}
X.-G. Wen {\em Adv. Phys.} {\bf 44} (1995) 405.

\bibitem{gaps}
L.~Georgiev, ``Stability and activation gaps of parafermionic {H}all states in
  the second {L}andau level,'' {\em Nucl. Phys.} {\bf B 626} (2002) 415--434,
  \href{http://xxx.lanl.gov/abs/cond-mat/0102451}{{\tt cond-mat/0102451}}.

\bibitem{cz}
A.~Cappelli and G.~R. Zemba, ``Modular invariant partition functions in the
  quantum {H}all effect,'' {\em \NP} {\bf B490} (1997) 595,
  \href{http://xxx.lanl.gov/abs/hep-th/9605127}{{\tt hep-th/9605127}}.

\bibitem{CFT-book}
P.~\uppercase{D}i Francesco, P.~Mathieu, and D.~S\'en\'echal, {\em Conformal
  Field Theory}.
\newblock Springer--Verlag, New York, 1997.

\bibitem{ctz3}
A.~Cappelli, C.~Trugenberger, and G.~Zemba, ``$w_{1+\infty}$ dynamics of edge
  excitations in the quantum {H}all effect,'' {\em Annals Phys.} {\bf 246}
  (1996) 86, \href{http://xxx.lanl.gov/abs/cond-mat/9407095}{{\tt
  cond-mat/9407095}}.

\bibitem{schw}
J.~Fuchs, B.~Schellekens, and C.~Schweigert, ``The resolution of field
  identification fixed points in diagonal coset theories,'' {\em \NP} {\bf
  B461} (1996) 371, \href{http://xxx.lanl.gov/abs/hep-th/9509105}{{\tt
  hep-th/9509105}}.

\bibitem{kt}
V.~Kac and I.~Todorov {\em Commun. Math Phys.} {\bf 190} (1997) 57--111.

\bibitem{JMP98}
L.~Georgiev and I.~Todorov, ``{RCFT} extensions of {$W_{1+\infty}$} in terms of
  bilocal fields,'' {\em J.~Math.~Phys.} {\bf 39} (1998) 5762--5771,
  \href{http://xxx.lanl.gov/abs/hep-th/9710134}{{\tt hep-th/9710134}}.

\bibitem{ctz}
A.~Cappelli, C.~A. Trugenberger, and G.~R. Zemba {\em \NP} {\bf B396} (1993)
  465.

\bibitem{laugh}
R.~Laughlin, ``Anomalous quantum {H}all effect: An incompressible quantum fluid
  with fractionally charged excitations,'' {\em Phys. Rev. Lett.} {\bf 50}
  (1983) 1395.

\bibitem{byers-yang}
N.~Byers and C.~Yang, ``Theoretical considerations concerning quantized
  magnetic flux in superconducting cylinders,'' {\em Phys. Rev. Lett.} {\bf 7}
  (1961) 46.

\bibitem{halpern-PF}
M.~Halpern {\em Ann. of Phys.} {\bf 194} (1989) 247.

\bibitem{halpern}
M.~Halpern, E.~Kiritsis, N.~Obers, and K.~Clubok, ``Irrational conformal field
  theory,'' {\em Phys. Rept.} {\bf 265} (1996) 1--138,
  \href{http://xxx.lanl.gov/abs/hep-th/9501144}{{\tt hep-th/9501144}}.

\bibitem{cristofano-PF}
G.~Cristofano, G.~Maiella, and V.~Marotta, ``A conformal field theory
  description of the paired and parafermionic states in the quantum {H}all
  effect,'' {\em Mod. Phys. Lett.} {\bf A 15} (2000) 1679.

\bibitem{gko}
P.~Goddard, A.~Kent, and D.~Olive {\em Commun. Math. Phys.} {\bf 103} (1986)
  105.

\bibitem{kac}
V.~Kac, {\em Infinite Dimensional Lie Algebras}.
\newblock Cambridge University Press, Cambridge, second~ed., 1985.

\bibitem{diag_coset}
L.~Georgiev, ``The diagonal affine coset construction of the ${\Z}_k$
  parafermion {H}all states,'' {\em proc. of the Fifth Int. Workshop "Lie
  Theory and its Applications in Physics", Eds. V. Dobrev et al., World
  Scientific, Singapore} (2004)
  \href{http://xxx.lanl.gov/abs/hep-th/0402159}{{\tt hep-th/0402159}}.

\bibitem{coleman}
S.~Coleman {\em Commun. Math. Phys.} {\bf 31} (1973) 259.

\bibitem{mermin-wagner}
N.~Mermin and H.~Wagner, ``Absence of feromagnetism in one- and two-
  dimensional isotropic {H}eisenberg models,'' {\em Phys. Rev. Lett.} {\bf 17}
  (1966) 1133.

\bibitem{ino2}
K.~Ino, ``Persistent edge currents for paired quantum {H}all states,'' {\em
  Phys. Rev.} {\bf B62} (2000) 6936,
  \href{http://xxx.lanl.gov/abs/cond-mat/0008094}{{\tt cond-mat/0008094}}.

\bibitem{kiryu}
H.~Kiryu, ``Oscillations of persistent edge currents in the parafermion quantum
  {H}all states,'' {\em Phys. Rev.} {\bf B65} (2002) 113407.

\bibitem{rr}
N.~Read and E.~Rezayi {\em \PR} {\bf B59} (1998) 8084.

\bibitem{CMP99}
A.~Cappelli, L.~Georgiev, and I.~Todorov, ``A unified conformal field theory
  approach to paired quantum {H}all states,'' {\em Commun. Math. Phys.} {\bf
  205} (1999) 657--689, \href{http://xxx.lanl.gov/abs/preprint ESI-621 (1998);
  hep-th/9810105}{{\tt preprint ESI-621 (1998); hep-th/9810105}}.

\bibitem{geller-loss-kircz-1}
M.~Geller, D.~Loss, and G.~Kirczenow, ``Luttinger liquids and composite
  fermions in nanostructures: what is the nature of the edge states in the
  fractional quantum {H}all effect?,'' {\em Superlattices Microstruct.} {\bf
  21} (1997) 5110.

\end{thebibliography}\endgroup

\end{document}